\documentclass{aa}
\usepackage{amsmath}
\usepackage{subfigure}
\usepackage[T1]{fontenc}
\usepackage{booktabs}
\usepackage{txfonts}
\usepackage{sistyle}
\usepackage{tablefootnote}
\usepackage{color}
\usepackage{graphicx}
\usepackage{mhchem}
\usepackage{natbib}

\title{Evolution of the reservoirs of volatiles in the protosolar nebula}

\author{Antoine Schneeberger \inst{1}, Olivier Mousis\inst{1}\inst{2}, Artyom Aguichine\inst{1}, \and Jonathan I. Lunine\inst{3}
}

\institute{Aix- Marseille Universit\'e, CNRS, CNES, Institut Origines, LAM, Marseille, France\\
              \email{antoine.schneeberger@lam.fr}
\and 
Institut Universitaire de France (IUF), France \\
\and
Cornell University, Department of Astronomy, Ithaca NY, USA\\
}

\date{Received August 02, 2022; accepted November 30, 2022}
\authorrunning{A. Schneeberger et al.}

\abstract{The supersolar abundances of volatiles observed in giant planets suggest that a compositional gradient was present at the time of their formation in the protosolar nebula. To explain this gradient, several studies have investigated the radial transport of trace species and the effect of icelines on the abundance profiles of solids and vapors formed in the disk. However, these models only consider the presence of solids in the forms of pure condensates or amorphous ice during the evolution of the protosolar nebula. They usually neglect the possible crystallization and destabilization of clathrates, along with the resulting interplay between the abundance of water and those of these crystalline forms. This study is aimed at pushing this kind of investigation further by considering all possible solid phases together in the protosolar nebula: pure condensates, amorphous ice, and clathrates. To this end, we used a one-dimensional (1D) protoplanetary disk model coupled with modules describing the evolution of trace species in the vapor phase, as well as the dynamics of dust and pebbles. Eleven key species are considered here, including H$_2$O, CO, CO$_2$, CH$_4$, H$_2$S, N$_2$, NH$_3$, Ar, Kr, Xe, and PH$_3$. Two sets of initial conditions are explored for the protosolar nebula. In a first scenario, the disk is initially filled with icy grains in the forms of pure condensates. In this case, we show that clathrates can crystallize and form enrichment peaks up to about ten times the initial abundances at their crystallization lines. In a second scenario, the volatiles were delivered to the protosolar nebula in the forms of amorphous grains.  In this case, the presence of clathrates is not possible because there is no available crystalline water ice in their formation region. Enrichment peaks of pure condensates also form beyond the snowline up to about seven times the initial abundances. Our model can then be used to compare the compositions of its different volatile reservoirs with those of comet C/2016 R2 PanSTARRS, Jupiter, Uranus, and Neptune. We find that the two investigated scenarios provide compositions of solids and vapors consistent with those observed in the bodies considered.}


\begin{document}

\maketitle
\section{Introduction}

It is commonly assumed that Solar System bodies have bulk compositions that are representative of the material present in the protosolar nebula (PSN) from which they formed. If the PSN was homogeneous in composition, gas and ice giants would be expected to reflect this homogeneity. However, observations show that giant planets present a range of supersolar metallicities. In Jupiter's atmosphere, the abundances of volatile elements were found to be $\sim$1.5--6.1 times higher than their protosolar values \citep{at03,mou18,li20}, with a few exceptions attributed to interior processes. In Uranus and Neptune, volatile abundances can reach up to about 100 times their protosolar values \citep{li87,li90,bai95,ka09,sr14}.

Numerical models show that the dynamics of icy pebbles and their vapors around icelines is an efficient mechanism to produce local changes in the composition of the PSN \citep{bo17,de17}. This process efficiently concentrates species around their respective icelines, creating the compositional gradient that may be responsible for the volatile enrichment in giant planets of our Solar System. The radial transport of trace species and the effect of icelines have been investigated using accretion disk models to assess the composition of the PSN. The resulting compositional profiles would then be used to constrain the formation conditions of gas giants \citep{mou19,sch21,ag22} or ice giants \citep{ow99,mon15,mo20}. This approach has also been used to explain the diversity among the cometary compositions and to determine their source or the general origins of their building blocks \citep{man20,mou21a}.

The form taken by volatiles that have fallen from the interstellar medium (ISM) onto the PSN is still an open question. Icy pebbles that are present at the earliest stages of the PSN may have formed in the very cold environment of the ISM, at temperatures of 10K or below \citep{gi04}. At such low temperatures, H$_2$O condenses in an amorphous structure that can efficiently trap other volatile species \citep{may86,je95}. At $\sim$135 K, the amorphous ice transitions to crystalline ice and releases trapped volatiles \citep{ba07}. This temperature is much higher than the usual sublimation temperature of pure condensates. In the PSN, the heliocentric distance at which this release occurs is called the Amorphous to Crystalline Transition Zone (ACTZ), and is located at approximately 5 au \citep{mou19}. However, it is not clear whether this amorphous ice survives the fall onto the PSN. For instance, it has been proposed that amorphous dust was heated up to crystallization when entering the PSN \citep{vi13}, as a consequence of the presence of shockwaves in the accreting PSN \citep{mi17,bu19}. As a result, many circumstellar disk models treat volatile species as pure condensates, with sublimation temperatures computed from thermodynamic tables \citep{dr09,ci15,ob19}. Alternatively, volatile species may also be trapped in clathrate form in the PSN, when sufficient amounts of H$_2$O are available \citep{lu85,ga05,mou21b}. A species, $i,$ is usually trapped in clathrate at a higher temperature than the one needed to sublimate its pure condensate form, except for the cases of CO$_2$ and NH$_3$. Because CO$_2$ clathrate and NH$_3$ monohydrate form at lower temperatures than their respective condensates at nebular pressures, they are not considered in our model. In the following, we refer to the heliocentric distance at which volatiles are entrapped in or released from clathrates as the clathration line. For this reason, volatile species can remain adsorbed on amorphous ice or trapped in clathrates, then they are released several au closer to the Sun -- than they otherwise would be if they were in the form of pure condensates.

\begin{figure*}[ht]

    \center
    \includegraphics[width= 1.1\textwidth, clip]{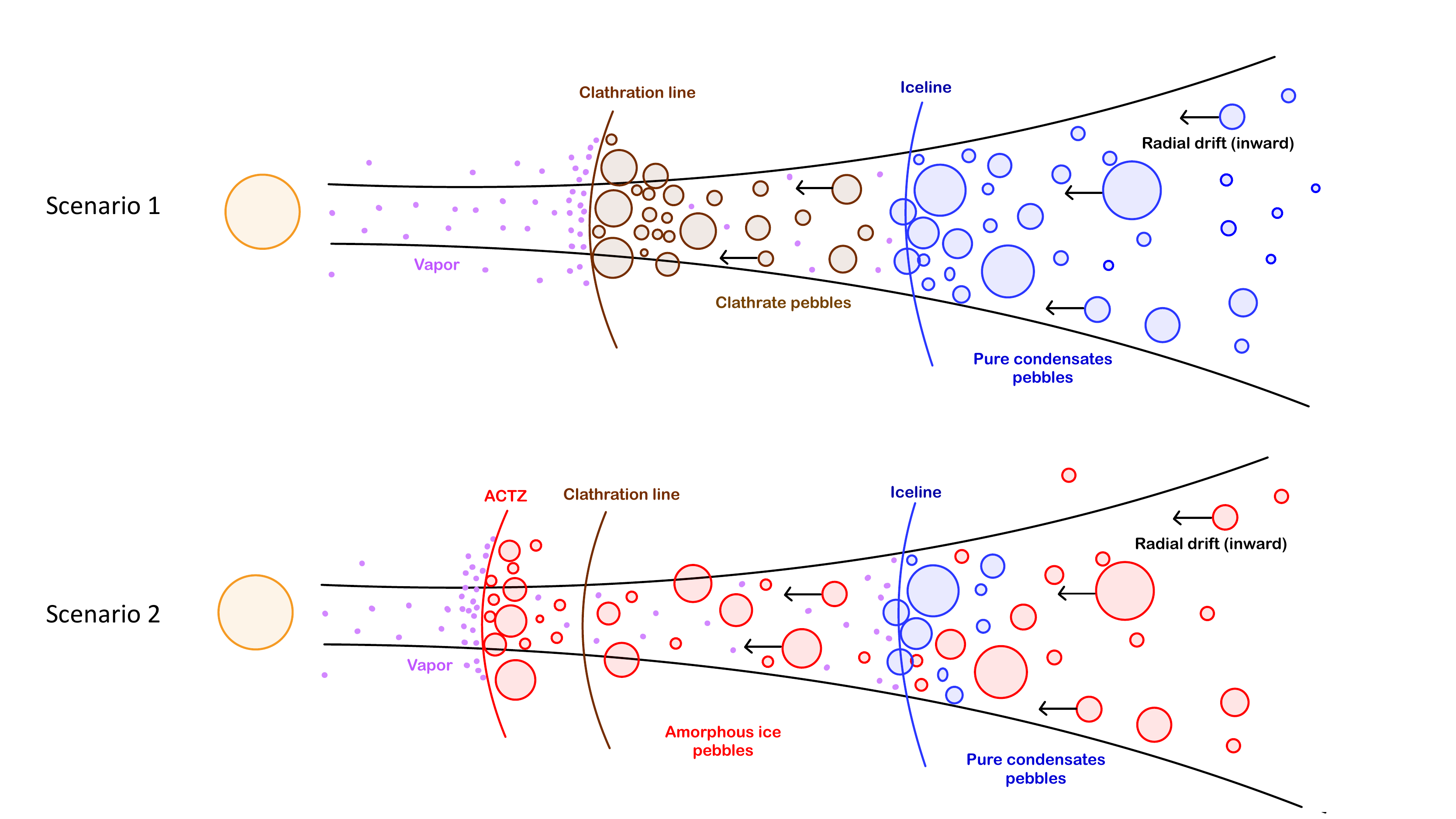}
    
    \caption{ Two outcome scenarios for volatile species explored in this paper. Top panel represents the case where volatiles are initially delivered in pure condensate form to the PSN (scenario 1). Bottom panel represents the case where volatiles are released in vapor form in the PSN when amorphous grains cross the ACTZ region (scenario 2). Pure condensates, clathrate, and amorphous ice pebbles are represented as blue, brown and red circles, respectively. Vapor is represented as purple dots. The iceline, clathration line, and ACTZ are represented as blue, brown, and red solid lines, respectively. Once delivered to the disk, the phase (solid or gaseous) of each species is determined by the positions of the corresponding condensation, hydration, or clathration lines. Except for the case of CO$_2$, which vaporises at a higher temperature than its clathrate form, hydration, or clathration lines of the volatiles considered are closer to the Sun than their respective icelines. Gaseous volatiles condense or become entrapped (depending on the availability of water ice) when diffusing outward of the locations of their condensation, hydration, or clathration lines. Conversely, volatiles condensed or entrapped in grains or pebbles are released in vapor form when drifting inward of their lines. Peaks of abundances form close to each phase-transition line (see text). Those enrichments are represented by higher solid and vapor concentrations in the panels.} 
    \label{fig:Scenarios}
\end{figure*}

In this study, we aim to quantify the influence of the presence of various icelines in the PSN, including the ACTZ and multiple clathration lines, on the nature of the main volatile reservoirs that were at play during the formation of the first icy grains in the disk. To do so, we use an existing PSN model that already describes the condensation and sublimation of pure ices, as well as the transport of species in solid and vapor forms \citep{ag20,ag22}, along with prescriptions of amorphous ice destabilization, as well as clathrate crystallization and dissociation added. Eleven key species are considered in our approach, namely: H$_2$O, CO, CO$_2$, CH$_4$, H$_2$S, N$_2$, NH$_3$, Ar, Kr, Xe, and PH$_3$. Each of these species can exist in the forms of vapor, crystalline ice, clathrate (monohydrate in the case of NH$_3$), or amorphous ice in the PSN. Two scenarios are investigated, each of them corresponding to a different initial state of the system. In scenario 1: the PSN is filled with volatile species in pure condensates or vapor form, depending on their location in the PSN. In scenario 2: the PSN is filled with volatile species adsorbed on amorphous ice beyond the ACTZ, and in vapor form in regions closer to the Sun. Because the ACTZ is located beyond the snowline, H$_2$O is found in crystalline form between the snowline and the ACTZ. Figure \ref{fig:Scenarios} represents the different forms of volatiles in the PSN in these two cases. In scenario 1, volatile species are in the vapor phase between the snowline and their respective icelines. If enough H$_2$O is available, these vapors can form clathrates. In scenario 2, icy grains of amorphous ice that drifted inward of the ACTZ release the adsorbed volatile species as vapors. The outward diffusion of these vapors can lead to the formation of clathrates or pure condensates (or both).

Section \ref{sec:sec2} describes the PSN model, as well as the transport modules used in our calculations. It also depicts the different formalisms that have been added to the disk model to mimic the formation and destabilization of clathrates, as well as the desorption of  volatiles from the amorphous ice particles crossing the ACTZ. In Section \ref{sec:sec3}, the radial profiles of the abundances of the different species are represented in various forms as a function of time in the PSN and in the individual cases described by the two scenarios. Section \ref{sec:sec4} is devoted to a discussion of the sensitivity of our results in light of the variation of the disk parameters. Our model is then used to compare the compositions of its resulting volatile reservoirs with those of the H$_2$O--poor comet C/2016 R2 PanSTARRS (R2), Jupiter, Uranus, and Neptune. This allows us to discuss the formation conditions of those bodies in the context of the two scenarios. Section \ref{sec:sec5} presents our conclusions.

\section{Volatile transport and evolution model}
\label{sec:sec2}

In this section, we describe the protosolar nebula model employed in our simulations, along with the modules calculating the transport of dust particles and vapors within the disk. Source and sink terms related to sublimation and condensation of pure ices as well as to clathrate destabilization and formation within the disk are also depicted.

\subsection{Protoplanetary disk model}

The disk model used here is the one described in \cite{ag20} and \cite{mo20}. The evolution of the PSN is governed by the following differential equation \citep{ly74}:

\begin{equation}
\frac{\partial \Sigma_{\mathrm{g}}}{\partial t} = \frac{3}{r} \frac{\partial}{\partial r} \left[ r^{1/2} \frac{\partial}{\partial r} \left( r^{1/2} \Sigma_{\mathrm{g}} \nu \right) \right],
\label{eq:psn}
\end{equation}

\noindent which describes the time evolution of a viscous accretion disk of surface density, $\Sigma_\mathrm{g}$, and viscosity, $\nu$, assuming invariance in the orbital direction and hydrostatic equilibrium in the azimuthal direction. This equation can be rewritten as a set of two first-order differential equations coupling the gas surface density, $\Sigma_{\mathrm{g}}$, field and mass accretion rate, $\dot{M}$:

\begin{equation}
\begin{cases}
\displaystyle \frac{\partial \Sigma_\mathrm{g}}{\partial t} = \frac{1}{2 \pi r} \frac{\partial \dot{M}}{\partial r} \\
\displaystyle \dot{M} = 3 \pi \Sigma_\mathrm{g} \nu \left( 1 + 2\frac{\partial \ln{\nu \Sigma_\mathrm{g}}}{\partial \ln{r}}\right)
\end{cases} .
\label{eq:resol_sys}
\end{equation}

\noindent  The first equation is a mass conservation law and the second one is a diffusion equation. The mass accretion rate is expressed as a function of the gas velocity field, $v_g$, and the radius, $r,$ as $\dot{M}=-2\pi  \Sigma_{\mathrm{g}} v_{\mathrm{g}} \mathbf{r} $.

The dynamical viscosity $\nu$ is calculated using the prescription of \cite{sh73}:

\begin{equation}
\nu = \alpha \frac{c_\mathrm{s}^2}{\Omega_{\mathrm{K}}},
\label{eq:nu}
\end{equation}

\noindent where $\alpha$ is the viscosity coefficient,  $c_\mathrm{s}$ is the sound speed in the PSN, and $\Omega_\mathrm{K}$ is the Keplerian frequency; also, $\alpha$ is estimated to be in the $10^{-4}$--$10^{-2}$ range, based on models calibrated on disk observations \citep{ha98,he04,ga05,bir12,dr17,ar19}. The sound speed, $c_\mathrm{s}$, is expressed as follows: 

\begin{equation}
c_s = \sqrt{\frac{R T}{\mu_\mathrm{g}}},
\label{eq:cs}
\end{equation}

\noindent where $\mu_\mathrm{g}$ is the mean molecular mass of the gas in the PSN, assumed here to be equal to \SI{2.31}{g.mol^{-1}}, $T$ is the midplane temperature, and $R$ is the ideal gas constant.

Two energy sources are considered in our model, namely viscous heating, and the constant irradiation by the local environment of ambient temperature, $T_{\mathrm{amb}}=$ \SI{10}{K}. Irradiation from the young Sun is neglected because the presence of shadowing is assumed in the outer part of the disk \citep{Oh21}. This allows the disk temperature to decrease to the condensation temperature of Ar ($\sim$20 K), allowing this species to be trapped in Jupiter's building blocks in a way consistent with the supersolar abundance observed in its envelope \citep{mou09b,mou12,mou21b}. The temperature profile is computed by summing the energy production rates of both energy sources \citep{hu05}:  

\begin{equation}
T^4 = \frac{1}{2\sigma_{\mathrm{SB}}} \left( \frac{3}{8}\tau_\mathrm{R} + \frac{1}{2 \tau_\mathrm{P}} \right) \Sigma_\mathrm{g} \nu \Omega_\mathrm{K}^2 + T^4_{\mathrm{amb}},
\label{eq:temp}
\end{equation}

\noindent where $\sigma_{\mathrm{sb}}$ is the Stefan-Boltzmann constant, while $\tau_\mathrm{R}$ and $\tau_\mathrm{P}$ are the Rosseland and Planck optical depth, respectively. Here, we assume  $\tau_\mathrm{P} = 2.4 \tau_\mathrm{R}$, a case corresponding to the opacity generated by dust grains smaller than 10 $\mu$m \citep{na94}; $\tau_\mathrm{R}$ is derived from the Rosseland mean opacity, $\kappa_\mathrm{R}$, via the following expression \citep{hu05}:

\begin{equation}
\tau_R = \frac{\Sigma_g \kappa_R}{2}.
\label{eq:kappaR}
\end{equation}

\noindent Here, $\kappa_\mathrm{R}$ is computed as a sequence of power laws of the form $\kappa_\mathrm{R} = \kappa_0 \rho^a T^b$, where $\rho$ denotes the gas density at the midplane, and $\kappa_0$, $a,$ and $b$ are constants that are obtained by fits on observational data in different opacity regimes \citep{be94}.

The initial state of the model is computed from the self-similar solution derived by \cite{ly74}: 

\begin{equation}
\Sigma_\mathrm{g} \nu \propto \exp{\left[ -\left( \frac{r}{r_\mathrm{c}} \right)^{2-p} \right]}.
\label{eq:selfsims}
\end{equation}

\begin{figure}[h]
\includegraphics[width= \columnwidth ]{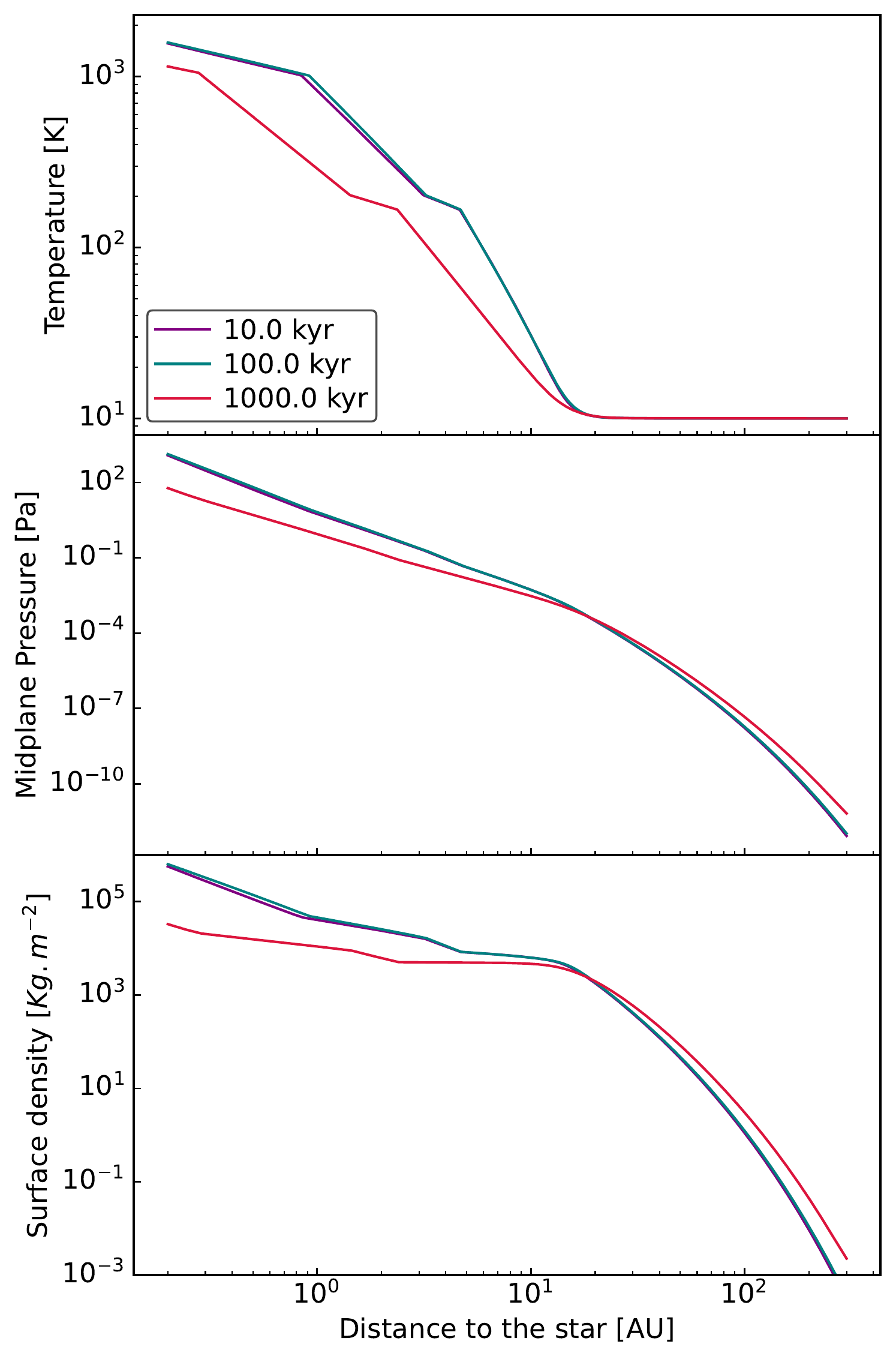}
\caption{Profiles of the disk midplane temperature, pressure and surface density calculated at $t$ = 10$^4$, 10$^5$, and 10$^6$ yr as a function of heliocentric distance, assuming $\alpha$ = $10^{-3}$, shown from top to bottom.}
\label{fig:PTSig}
\end{figure}

\noindent By combining Eqs. \ref{eq:selfsims} and \ref{eq:resol_sys}, and assuming $p=\frac{3}{2}$, which corresponds to the case for an early disk \citep{ly74}, the initial profiles of the dust surface density and mass accretion rate are given by:

\begin{equation}
\begin{cases}
\displaystyle \Sigma_{\mathrm{g},0} = \frac{\dot{M}_{\mathrm{acc},0}}{3 \pi \nu } \exp \left[ - \left( \frac{r}{r_\mathrm{c}} \right)^{0.5} \right] \\
\displaystyle \dot{M}_0 = \dot{M}_{\mathrm{acc},0} \left(  1 - \left( \frac{r}{r_\mathrm{c}} \right)^{0.5}\right) \exp\left[- \left( \frac{r}{r_\mathrm{c}} \right)^{0.5} \right] 
\end{cases},
\label{eq:init}
\end{equation}

\noindent where $r_\mathrm{c}$ is the centrifugal radius and $\dot{M}_{\mathrm{acc},0}$ is the initial mass accretion rate onto the central star, set to \SI{10^{-7.6}}{M_{\odot}.yr^{-1}} \citep{ha98}. The disk mass is related to the surface density profile via the following expression:

\begin{equation}
M_{\mathrm{disk}} = 2\pi \int_{R_{\min}}^{R_{\max}} \Sigma r dr,\label{eq:init_mass}
\end{equation}

where $R_{\min}$ and $R_{\max}$ are the inner and outer bounds of our model. The total disk mass is set to $0.1M_{\odot}$ and most of it (99\%) is encapsulated within $\sim$200 au. The centrifugal radius, $r_\mathrm{c}$, is determined by solving Eqs. \ref{eq:init} with \ref{eq:init_mass} for the chosen values of mass accretion rate and disk mass. Figure \ref{fig:PTSig} represents the thermodynamic profiles of our PSN model assuming $\alpha$ = $10^{-3}$, and at $t$ = 10$^4$, 10$^5$, and 10$^6$ yr of the disk evolution.

\subsection{Dust dynamics}
\label{sec2.2}

To determine the size of the dust pebbles, we rely heavily on the two-population algorithm developed by \cite{bir12}. This algorithm relies on the key idea that the dynamics of dust pebbles of many different sizes can be well approximated by the dynamics of only two populations of particles, in which all particles have the same representative sizes. The first group corresponds to the small population, where grains are of constant size: $a_0 =$ 0.1 $\mu$m. The second group represents a large population, where pebbles have a representative size $a_1$, which depends on the characteristics of the flow.

In the disk, pebbles grow by sticking collisions via the following law:
\begin{equation}
a_1(t) = a_0 \exp{\left( \frac{t}{\tau_{\mathrm{growth}}} \right)},
\end{equation}
where $\tau_{\mathrm{growth}}$ is the growth timescale,
\begin{equation}
 \tau_{\mathrm{grow}} = \frac{4 \Sigma_\mathrm{g}}{\sqrt{3}\epsilon_\mathrm{g} \Sigma_\mathrm{b} \Omega_\mathrm{K}},
\end{equation}
where $\Sigma_{\mathrm{b}}$ is the total surface density of solids, and $\epsilon_\mathrm{g}$ is the dust growth efficiency through mutual sticking set to 0.5 \citep{la14}. Then, we compute the Stokes number of pebbles as a function of their sizes \citep{jo14}:
\begin{equation}
\mathrm{St} = 
\begin{cases}
\displaystyle \sqrt{2 \pi} \frac{a_1 \rho_\mathrm{b}}{\Sigma_\mathrm{g}} \text{    If      } a_1 \leq \frac{9}{4} \lambda \\ 
\displaystyle \frac{8}{9} \frac{a_1^2 \rho_b c_\mathrm{s}}{\Sigma_\mathrm{g} \nu} \text{    If      } a_1 \geq \frac{9}{4} \lambda
\end{cases}.
\label{eq:stokes}
\end{equation}
The top and bottom lines of Eq. \ref{eq:stokes} correspond to the Epstein and Stokes regimes, respectively. The limit between both regimes is fixed by the gas mean free path $\lambda = \sqrt{\pi/2}  \cdot \nu / c_\mathrm{s}$, computed by equating the two terms in Eq.(\ref{eq:stokes}). The term $\rho_{\mathrm{b}}$ is the pebbles mean bulk density:
\begin{equation}
\rho_{\mathrm{b}} = \frac {\sum_i \Sigma_{\mathrm{b},i} \rho_{\mathrm{b},i} } { \sum_i \Sigma_{\mathrm{b},i} },
\end{equation}
computed as the average of each species' bulk density, $\rho_{\mathrm{b},i}$, weighted by their solid surface density, $\Sigma_{\mathrm{b},i}$.

Observations indicate that disks are rich in small dust and suggest that fragmentation is a dominant process \citep{wi11}. Based on this observation, our approach considers fragmentation and radial drift as the growth-limiting mechanisms. These mechanisms set an upper limit on the highest Stokes number that particles can achieve. The first limitation results from the fragmentation occurring when the relative speed between two pebbles due to turbulent motion exceeds the fragmentation velocity, $u_\mathrm{f}$. This upper limit is given by \citep{bir12}: 

\begin{equation}
\mathrm{St}_{\mathrm{frag}} = f_{\mathrm{f}} \frac{1}{3 \alpha} \frac{u_\mathrm{f}^2}{c_\mathrm{s}^2},
\end{equation}

\noindent where $u_\mathrm{f}$ is set to $\SI{10}{m.s^{-1}}$ and the factor $f_{\mathrm{f}}=0.37$ accounts for the fact that the representative size of the large population is smaller than the biggest size particles can achieve before they fragment. 

A second limitation for dust growth is determined by the drift velocities of the different pebbles. When pebbles drift faster than they grow, this sets another upper limit for the Stokes number \citep{bir12}: 

\begin{equation}
\mathrm{St}_{\mathrm{drift}} = f_{\mathrm{d}} \frac{\Sigma_\mathrm{b} v_\mathrm{K}^2}{\Sigma_\mathrm{g} c_\mathrm{s}^2} \left| \frac{\mathrm{d} \ln P}{\mathrm{d} \ln r} \right|^{-1},
\label{eq:stdrift}
\end{equation}

\noindent where $P$ is the disk midplane pressure, $v_\mathrm{K}$ the keplerian velocity, and $f_{\mathrm{d}}=0.55$ has the same origin as $f_\mathrm{f}$.

When dust grains drift at a high velocity and collide with other particles on their path, they can fragment. This induces a third upper limit for the Stokes number \citep{bir12}, given by: 
\begin{equation}
\mathrm{St}_\mathrm{df} =\frac{1}{1-N} \frac{u_{\mathrm{f}} v_{\mathrm{K}}}{c_{\mathrm{s}}} \left( \frac{\mathrm{d} P}{\mathrm{d} r} \right)^{-1},
\end{equation}

\noindent where the factor $N=0.5$ accounts for the fact that only larger grains fragment when colliding.

In the algorithm, all limiting Stokes numbers are computed and compared with the Stokes number derived from Eq. \ref{eq:stokes}. At each time step, the smallest Stokes number found in this comparison becomes the reference Stokes number which, in turn, sets the value for the representative size, $a_1$, of the large population. The representative size of the small population is always $a_0$, and their Stokes number is always computed in the Epstein regime (top line of Eq. \ref{eq:stokes}). 

Finally, the two-population algorithm of \cite{bir12} introduces $f_\mathrm{m}$ the fraction of the mass contained in the large population. Among the three size-limiting mechanisms, if particle drift is the most limiting one ($\mathrm{St}_\mathrm{drift} = \min \left(\mathrm{St}_\mathrm{frag},\mathrm{St}_\mathrm{drift},\mathrm{St}_\mathrm{df}\right)$), then the fraction of the mass contained in the large population is $f_\mathrm{m} = 0.97$. Otherwise, $f_\mathrm{m}$ is set to 0.75 \citep{bir12}. The mean grain size $\bar{a}$ is then given by: 

\begin{equation}
\bar{a} = f_\mathrm{m} a_1 + (1 - f_\mathrm{m}) a_0.
\end{equation}

\subsection{Trace species evolution model}
\label{sec:Trace species evolution model}

Trace species are considered in four distinct forms: vapors, pure condensates, entrapped in clathrates, or forming a monohydrate (case for NH$_3$ only), and adsorbed in amorphous ice. In our model, a distinct surface density is attributed to each of these forms, with $\Sigma_{\mathrm{v},i}$, $\Sigma_{\mathrm{p},i}$, $\Sigma_{\mathrm{c},i}$, and $\Sigma_{\mathrm{a},i}$ corresponding to species $i$ in vapor, pure condensate, clathrate or hydrate, or amorphous ice phases, respectively. Their time and radial evolution is governed by the advection-diffusion equation \citep{bir12,de17}: 
\begin{equation}
\frac{\partial \Sigma_i}{\partial t} + \frac{1}{r} \frac{\partial}{\partial r} \left[ r \left( \Sigma_i v_i - D_i \Sigma_\mathrm{g} \frac{\partial}{\partial r} \left( \frac{\Sigma_i}{\Sigma_\mathrm{g}} \right) \right) \right] - \dot{Q}_i = 0,
\label{eq:advdiff}
\end{equation}
where $D_i$ is the diffusion coefficient and is $v_i$ is the radial speed. $\dot{Q}_i$ is a source or sink term that accounts for phase changes, counted positive/negative when some matter is created or lost.

For surface densities of vapors, we assume $D_i$ = $D_\mathrm{g}$ and $v_i$ = $v_\mathrm{g}$ because vapors are well coupled to the PSN gas and evolve similarly. The gas diffusivity, $D_\mathrm{g}$, is assumed to be equal to the viscosity, $\nu$, and the gas velocity is \citep{sh73}:
\begin{equation}
v_\mathrm{g} = - \frac{ \dot{M}_{\mathrm{acc}} }{2 \pi r \Sigma_\mathrm{g}}.
\end{equation}

At each time and location, we assume that dust particles are formed from a mixture of all available solids. As a consequence, surface densities of solid phases, namely clathrates (and NH$_3$ monohydrate), amorphous ices, and pure condensates, are evolved with the same diffusion coefficient, $D_\mathrm{s}$, and radial velocity, $v_\mathrm{s}$. For particles of a given size, $a,$ and Stokes number, St, the diffusion coefficient is given by \citep{bir12}:
\begin{equation}
D_\mathrm{s} = \frac{D_\mathrm{g}}{1 + \mathrm{St}^2}.
\end{equation}

\noindent The dust radial velocity is expressed as the sum of gas drag and drift velocities \citep{bir12}:

\begin{equation}
v_\mathrm{s} = \frac{1}{1 + \mathrm{St}^2}v_\mathrm{g} + \frac{2 \mathrm{St}}{1 + \mathrm{St}^2}v_{\mathrm{drift}},
\end{equation}
where the drift velocity is \citep{we97}:
\begin{equation}
v_{\mathrm{drift}} = \frac{c_\mathrm{s}^2}{v_\mathrm{K}} \frac{\mathrm{d}\ln P}{\mathrm{d} \ln r}.
\end{equation}
 
\noindent The diffusion coefficient and radial velocity of solids are computed for the small and the large populations, that is, for particles of sizes $a_0$ and $a_1$. The diffusion coefficient, $D_{\mathrm{s}}$, and radial velocity $v_{\mathrm{s}}$ used to evolve surface densities of solids are then given by mass-averaged diffusivities and velocities of the small and large population \citep{bir12}:

\begin{equation}
\begin{cases}
\displaystyle v_{\mathrm{s}}= f_\mathrm{m} v_{\mathrm{d},a_1} + (1-f_\mathrm{m}) v_{\mathrm{d},a_0}.\\
\displaystyle D_{\mathrm{s}}= f_\mathrm{m} D_{\mathrm{d},a_1} + (1-f_\mathrm{m}) D_{\mathrm{d},a_0}.
\end{cases}
\end{equation}

\subsection{Sources and sinks of trace species}

We follow the approach of \cite{ag20} to depict the sources and sinks for both the solid and vapor phases of the different species. A pure condensate of species $i$ undergoes sublimation if its partial pressure is lower than the corresponding equilibrium pressure. Sublimation results in a sink term for pure condensates during the time step, $\Delta t$ \citep{dr17}: 

\begin{equation}
\Dot{Q}_{\mathrm{p},i} =  - \min \left( \sqrt{\frac{8 \pi \mu_i}{RT}} \frac{3}{\pi \bar{a} \bar{\rho}} P_{\mathrm{eq},i} \Sigma_{\mathrm{p},i} ; \frac{\Sigma_{\mathrm{p},i}}{\Delta t} \right),
\label{eq:ev}
\end{equation}

\noindent where $\mu_i$ is the molar mass of species $i$, $\bar{\rho}$ is the mean bulk density of grains, $P_{\mathrm{eq},i} $ is the equilibrium pressure, and $\bar{a}$ is the mean size of grains. The second part of the minimum function ensures that no more than the available quantity of pure condensate sublimates. Equilibrium curves of pure condensates are given in Appendix \ref{an:condensates}. 

Conversely, a gas of species $i$ forms a pure condensate if its partial pressure is larger than the corresponding equilibrium pressure. The condensation rates results in a source term for pure condensates \citep{dr17}: 

\begin{equation}
\Dot{Q}_{\mathrm{p},i} = \min{\left( \left(P_i - P_{\mathrm{eq},i} \right) \frac{2H\mu_i}{RT \Delta t} ; \frac{\Sigma_{\mathrm{v},i}}{\Delta t} \right) }. 
\label{eq:cond}
\end{equation}

\begin{figure}[h]
\center
\includegraphics[width=0.9\columnwidth]{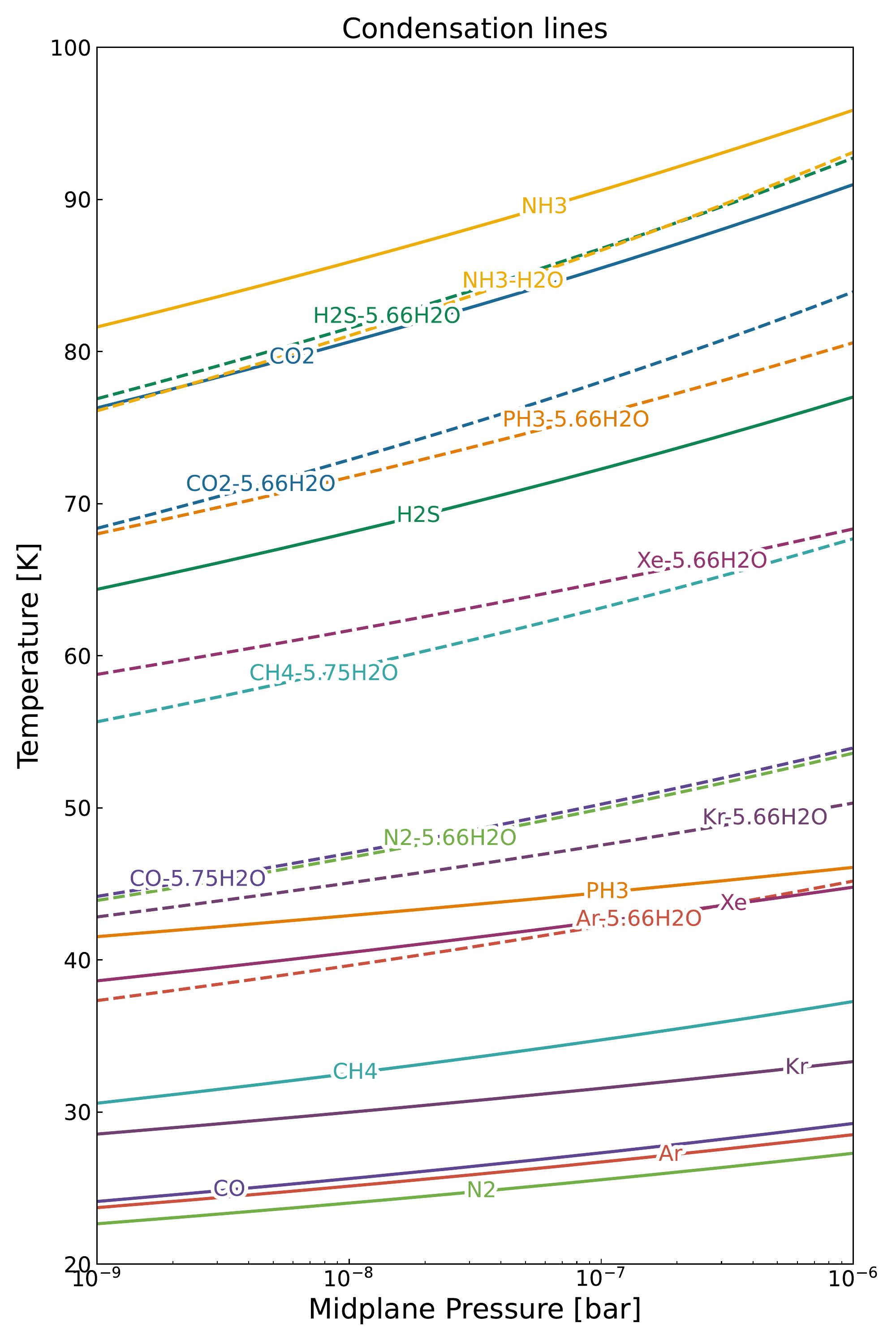}
\caption{Equilibrium curves of pure condensates (solid lines), clathrates, and NH$_3$ monohydrate (dashed lines) in a pressure-temperature domain relevant to PSN conditions (see the appendix for the relevant data). Two clathrate structures are considered in our model, namely, type I and type II with stoichiometric factors of 5.75 and 5.66, respectively. Partial pressures are calculated by considering the species abundances given in Table \protect{\ref{tab:Abundance}}.}
\label{fig:PvsT}
\end{figure}

We also added the possibility of NH$_3$ monohydrate and clathrate crystallization in the PSN. Assuming that enough crystalline water is available, these solids form first during the cooling of the disk because their crystallization temperatures are higher than those of the corresponding pure condensates (see Fig. \ref{fig:PvsT}). The only exceptions to that rule are CO$_2$  and NH$_3$, namely the only species that condense at a higher temperature than their hydrates at nebular conditions (see Fig. \ref{fig:PvsT}). To compute the source and sink terms of trace species in clathrates, we used the same prescription as that used for pure condensates (Eqs. \ref{eq:ev} and \ref{eq:cond}). The equilibrium pressures of pure condensates are replaced by those of clathrates and NH$_3$ monohydrate (see Appendix \ref{an:clathrate}). The formation of clathrates and NH$_3$ monohydrate also requires the presence of a minimum amount of crystalline water, resulting in a limit for their source term $\Dot{Q}_{\mathrm{c},i}$:

\begin{equation}
\Dot{Q}_{\mathrm{c},i} \Delta t \le \frac{\mu_{i}}{S_i \mu_{\ce{H2O}}} \left[ \Sigma_{\mathrm{p},\ce{H2O}} - \sum_k \Sigma_{\mathrm{c},k} \frac{S_k \mu_{\ce{H2O}}}{\mu_{k}} \right] 
\label{eq:cond_cla}
.\end{equation}

This expression takes the amount of available crystalline water $\Sigma_{\mathrm{p,\ce{H2O}}}$ and subtracts the amount of water that is already used to trap currently existing clathrates. In this expression, $S_k$ is the stoichiometric ratio between the species $k$ and water. This ratio is set to 5.75, 5.66, and 1 in the cases of type I clathrate, type II clathrate, and NH$_3$ monohydrate, respectively. This condition sets an upper limit on the clathrate source term, that can be equal to 0 if all the crystalline water is already holding clathrates. If all conditions are met for a trace species to be able to form pure condensates, clathrates, and monohydrates, we prioritize the solid phase that has the greatest $P_i - P_{\mathrm{eq,i}}$ value. Prioritizing the largest $P_i - P_{\mathrm{eq,i}}$ value is equivalent to prioritizing the highest solidification rate. Such considerations are not taken into account when performing a prioritization test among clathrate formation and condensation of pure condensates, since clathrates only form when pure condensates are not stable.

When grains containing amorphous ice cross the ACTZ, water crystallizes and releases all the adsorbed volatiles irreversibly. The corresponding sink term is then:

\begin{equation}
\dot{Q}_\mathrm{a} = - \frac{\Sigma_\mathrm{a}}{\Delta t},
\label{eq:am}
\end{equation}

\noindent with the only condition that must be satisfied is: $T > T_{\mathrm{ACTZ}}$.

From the various rates of condensation, crystallization and sublimation, we can finally derive the overall sink and source term for the vapor:

\begin{equation}
\Dot{Q}_\mathrm{v, \it i} = -\Dot{Q}_{\mathrm{p,\it i}} - \Dot{Q}_{\mathrm{c,\it i}} - \Dot{Q}_{\mathrm{a, \it i}}.
\end{equation}

\section{Results}
\label{sec:sec3}

Simulations have been performed in the case of two distinct scenarios. In scenario 1, particles initially released in the PSN are only made from pure condensates. During their inward drift, they can sublimate, condense again and/or form various clathrates and a NH$_3$ monohydrate, depending on the local temperature and pressure conditions of the disk. In scenario 2, particles initially released in the PSN are only made from amorphous ice. During their inward drift, water contained in these particles transitions to a crystalline structure when crossing the ACTZ and leads to the release of adsorbed volatiles in the gas phase. The released vapors can in turn condense into pure ices and/or form various clathrates and NH$_3$ monohydrate, following the prescription described in the previous section. Both scenarios are explored with the assumption of $0.1 M_{\odot}$ for the mass of the PSN and a viscosity parameter of $\alpha = 10^{-3}$.

The initial surface density of a species, $i,$ is:
\begin{equation}
\Sigma_{0,i} = x_{0,i} \Sigma_\mathrm{g},
\end{equation}
where $x_{0,i}$ is the initial mass fraction of the species, $i$. At $t=0$, the partial pressures of the different species are computed at each point of the grid. In scenario 1, if the partial pressure of a given species is below its equilibrium pressure, then the corresponding location is filled with vapor. Otherwise, this location filled with pure ice. In scenario 2, species are in the vapor phase where $T\ge T_\mathrm{ACTZ}$ and in an amorphous ice phrase otherwise. In both scenarios, thermodynamic equilibrium is established after a few time steps ($\sim1$ yr).

The initial PSN composition is derived from the protosolar elemental abundances tabulated by \cite{lod09}. We assume that all \ce{C} is distributed between \ce{CO}, \ce{CO2}, or \ce{CH4}, with the remaining \ce{O} forming \ce{H2O}. We have set \ce{CO}:\ce{CO2}:\ce{CH4} = 10:4:1 in the PSN gas phase. The \ce{CO}:\ce{CO2} ratio is derived from ROSINA measurements of comet 67P/C-G between 2014 August and 2016 September \citep{mou14}. The \ce{CO}:\ce{CH4} ratio is consistent with the production rates measured in the southern hemisphere of the 67P/C-G in October 2014 by the ROSINA instrument \citep{ro15}. Sulfur is assumed to be half in \ce{H2S} form and half in refractory sulfide components \citep{pa05}. We also assumed \ce{N2}:\ce{NH3} = 1:1, a value predicted by thermochemical models that take into account catalytic effects of Fe grains on the kinetics of N$_2$ to NH$_3$ conversion of the PSN \citep{Fe00,mou09a}. The molar abundances of the different species are derived from the gas phase abundances given in Table \ref{tab:Abundance}.

\begin{table}[h]
\centering
\caption{Initial molar abundances of the considered trace species.}
\begin{tabular}{@{}cccc@{}}
\hline
\hline           
\smallskip
Trace species   &  (X/H$_2$)$_\odot$    & Trace species &  (X/H$_2$)$_\odot$ \\
\hline           
\smallskip
\ce{H_2O}       & $5.479 \times 10^{-4}$        &       \ce{NH_3}       & $5.456 \times 10^{-5} $         \\
\ce{CO}         & $3.698 \times 10^{-4}$        &       \ce{PH_3}       & $6.368 \times 10^{-7} $ \\
\ce{CO_2}       & $1.479 \times 10^{-4}$        &       \ce{Ar}         & $7.150 \times 10^{-6} $         \\
\ce{CH_4}       & $3.698 \times 10^{-5}$        &       \ce{Kr}         & $4.310\times 10^{-9} $  \\
\ce{H_2S}       & $1.633 \times 10^{-5} $       &       \ce{Xe}         & $4.210 \times 10^{-10} $        \\
\ce{N_2}        & $5.456 \times 10^{-5} $       &                               &                                                       \\
\hline           
\end{tabular}
\label{tab:Abundance}
\end{table}

\begin{figure*}[ht]
\center
\includegraphics[width=\linewidth]{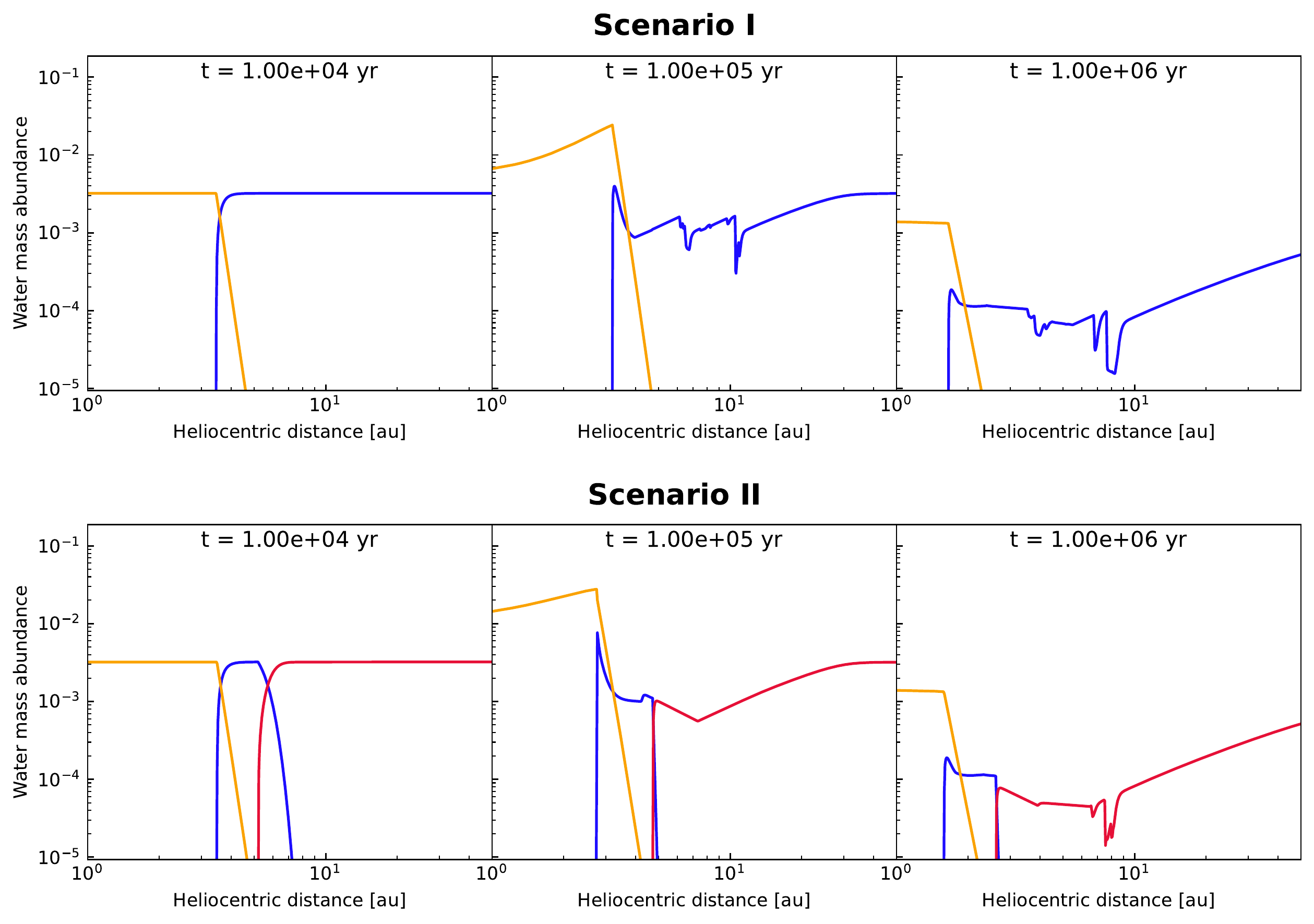}
\caption{Time evolution of the mass abundance of water, defined as the radial profile of $\Sigma_{\ce{H2O}} / \Sigma_\mathrm{g}$, for both scenarios at t = 10$^4$, 10$^5$, and 10$^6$ yr. Solid lines represent \ce{H2O} in the gaseous phase (orange line), crystalline phase (blue line), and in amorphous phase (red line).} 
\label{fig:water}
\end{figure*}

Figure \ref{fig:water} represents the radial evolution of the water mass abundance in the PSN at different epochs of its evolution in scenario 1 and  scenario 2 (top row and bottom row, respectively). Both cases produce very similar water abundances and show a water enrichment peak that is about 10 times its initial abundance at the location of the snowline ($\sim$2.8 au). However, the availability of crystalline ice is much more limited in scenario 2 due to \ce{H2O} being mostly in an amorphous state. Indeed, in this case, crystalline ice is present only in the $\sim$2.8--4 au region. 

\subsection{Scenario 1: Initial delivery of pure condensates}
\label{sec:Scenario I}

\begin{figure*}[ht]
\centering
\includegraphics[width = 0.76\linewidth]{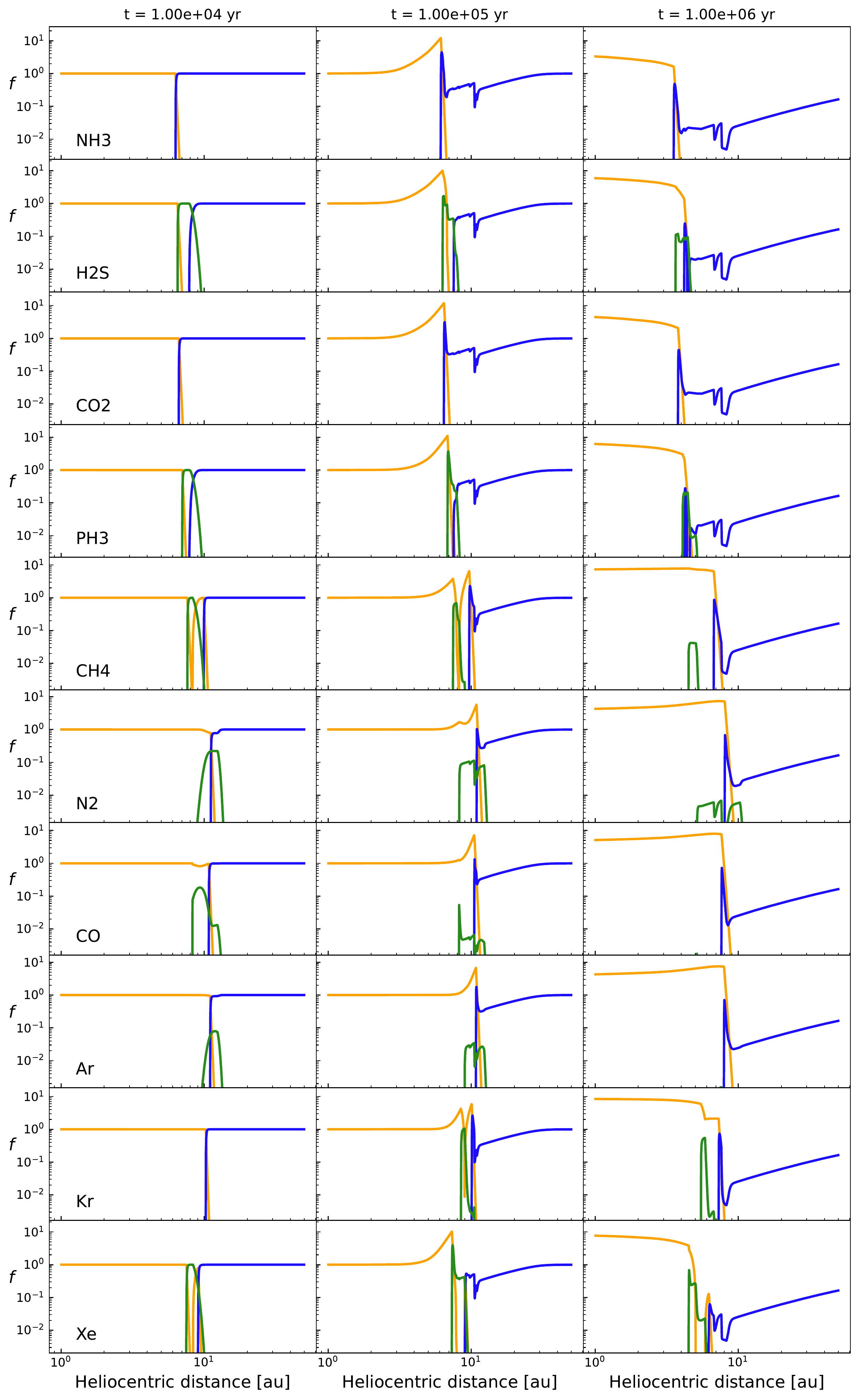}
\caption{Time and radial evolution of species' mass abundances normalized to their initial values in gaseous phase (orange line), pure condensate form (blue line), and clathrate (green line), at t = 10$^4$, 10$^5$, and 10$^6$ yr in scenario 1.}
\label{fig:cryst1-f}
\end{figure*}

Figure \ref{fig:cryst1-f} represents the time and radial evolution of the abundance ratios (with respect to initial abundances) of the different species existing in various phases in the case of scenario 1, respectively. After 10 kyr of PSN evolution, the species considered are present under all their possible forms -- except for CO$_2$ and NH$_3$, which are never enclathrated. By successive order with progressing heliocentric distance (and decreasing temperature and pressure conditions), we first find  the different vapors, then narrow regions corresponding to the presence of clathrates, and, finally,  an outer region that is only populated with pure condensates. Regions where clathrates exist expand from 7 to 12 au, depending on the species considered. At this early stage of the PSN evolution, there is no significant enrichment that can be observed for any species.

After 0.1 Myr of disk evolution, Figure \ref{fig:cryst1-f} also shows that beyond 6 au, clathrates coexist with pure condensates and their abundances decrease steeply -- except for CO$_2$ and NH$_3$, which that exist only as pure ices. A few au beyond that, pure condensates become the only solid structures existing in the outer PSN. Depending on the PSN ther	modynamic conditions and the availability of crystalline water, when the different species become fully enclathrated, their abundances form unique enrichment peaks located at their clathration lines, reaching up to $\sim$15 times their initial values. On the other hand, if the budget of crystalline water is not high enough, then the species are only partly enclathrated. They then form two narrow enrichment peaks at their condensation and clathration lines, reaching $\sim$5 and $\sim$15 times their initial abundances, respectively. Table \ref{tab:peaks_cryst} displays the heliocentric distance and the value of the enrichment peak (relative to the initial abundance) for each species under consideration at $t$ = 0.1 Myr of the PSN evolution. Depending on the species considered, the enrichment peaks range between 2 and 18 times the protosolar values and are located in the  2--11 au region. The closest and furthest peaks from the Sun are those of the water snow line at 2.8 au and N$_2$ iceline at 10.8 au, respectively. Two enrichment peaks, located at the clathration and icelines, are found in the cases of CH$_4$ and Kr. 

In Figure \ref{fig:cryst1-f}, all species, except CO, N$_2$, and Ar, exhibit a dip in the surface densities of the pebbles of pure condensates around 10.5 au. In this region, the total surface density of icy pebbles is increased due to the combined actions of CO, N$_2$, and Ar icelines located at 10.4, 10.9, and 10.7 au, respectively. The condensations rates of CO, N$_2$, and Ar locally increase the surface density of solids, which are in excess compared with their loss rates via inward drift. On the other hand, because the icelines of the other species are located closer to the Sun, their surface densities progressively decrease in this region, as a result of the inward  drift of particles.

After \SI{1}{Myr} of PSN evolution, all peaks are smoothed out in the gas phase. Because of the inward drift of pebbles, most of the species are in vapor form when the solids start to deplete. The total mass of solids has decreased by a factor of four compared to the beginning of the simulation. This effect is more significant in the case of clathrates forming at very low temperatures (less than 50 K), since the fraction of crystalline water used to form higher temperature clathrates increases with time.

\begin{table}[h]
\centering
\caption{Heliocentric distance and value of the enrichment peak (relative to the initial abundance) for each species under consideration at $t$ = 0.1 Myr in the case of scenario 1 (middle column of Figure \ref{fig:cryst1-f}).}
\begin{tabular}{@{}ccc@{}}
\hline
\hline
Element                 & Peak location (au)            & Peak value \\
\hline
\ce{H2O}                & 2.7                                   & 11.0          \\
\ce{CO}                 & 10.4                                  & 8.1           \\
\ce{CO_2}               & 6.0                                   & 16.0          \\
\ce{CH_4}               & 7.1                                   & 4.7           \\
                & 9.5                                   & 9.4           \\
\ce{H_2S}               & 5.9                                   & 14.0          \\
\ce{N_2}                & 10.9                                  & 5.6           \\
\ce{NH_3}               & 5.6                                   & 17.2          \\
\ce{Ar}                 & 10.7                                  & 7.3           \\
\ce{Kr}                 & 8.2                                   & 9.12          \\
                & 9.8                                   & 23.5          \\
\ce{Xe}                 & 7.0                                   & 14.3          \\
\ce{PH_3}               & 6.4                                   & 15.4          \\ 
\hline
\end{tabular}\\
Both vapor and solid forms are considered.
\label{tab:peaks_cryst}
\end{table}

\subsection{Scenario 2: initial delivery of amorphous ices} 
\label{sec:Scenario II} 

\begin{figure*}[ht]
\centering
\includegraphics[width = 0.76\linewidth]{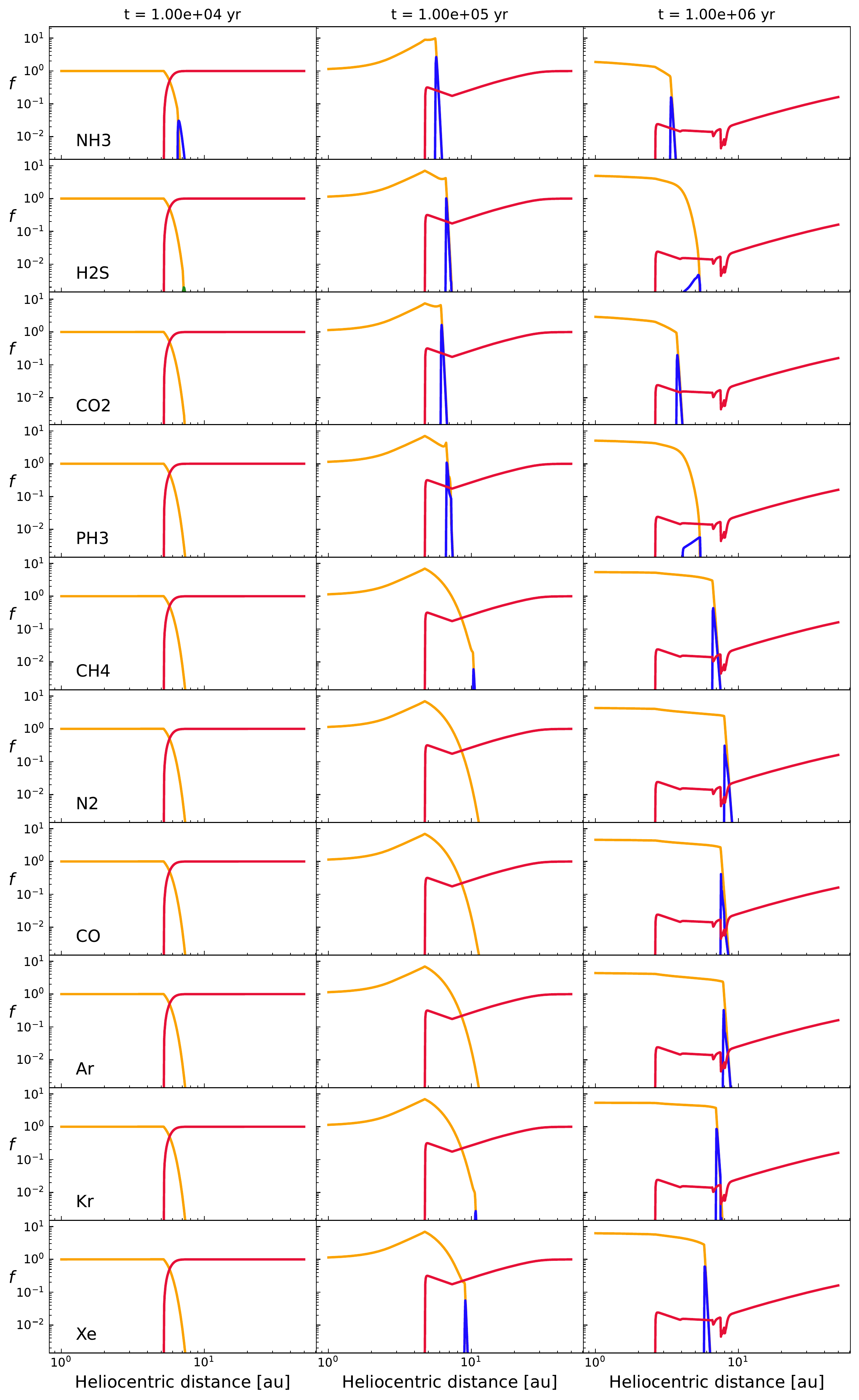}
\caption{Time and radial evolution of species' mass abundances normalized to their initial values in gaseous phase (orange line), pure condensate form (blue line), and amorphous form (red line), compared with their initial mass abundances, at t = 10$^4$, 10$^5$, and 10$^6$ yr in scenario 2.}
\label{fig:am1-f}
\end{figure*}

Figure \ref{fig:am1-f} represents the time and radial evolutions of the abundance ratios (with respect to the initial abundances) of the different species existing in various phases in the case of scenario 2. After 10 kyr of PSN evolution, with progressing heliocentric distance, first the different vapors are found, then an outer region that is only populated with amorphous ice. The budget of solid phase is dominated by species trapped in amorphous ice. Pure condensates only form when the vapors desorb from amorphous ice at the ACTZ (shown at $\sim$5 au), diffuse outward and cross an iceline. Such a process is efficient only if the condensation line is close to the ACTZ, leading to very narrow regions with the presence of pure condensates only in the case of NH$_3$.

After 0.1 Myr of disk evolution, it is notable that only pure condensates of NH$_3$, H$_2$S, CO$_2$, and PH$_3$ form. Table \ref{tab:peaks_am} displays the heliocentric distance and the value of the enrichment peak (relative to the initial abundance) for each species under consideration at this epoch of the PSN evolution. All peaks are located in a much narrower region, centered at the location of the ACTZ, compared with the scenario 1, with values ranging from 7 to 10 times the initial abundances. The peak locations are influenced by the presence of narrow (less than 1 au) regions filled with pure condensates. In those regions, corresponding to the icelines, the abundance of pure condensates exceed that of amorphous ice, thus influencing the location of the enrichment peaks.

After \SI{1}{Myr} of PSN evolution, as a result of pebble drift, all peaks have been smoothed and the volatiles trapped in amorphous water ice phase are strongly depleted by factors reaching more than 100, compared with their gaseous abundances in the inner disk. The amount of pure condensates exceeds that of amorphous ice in some narrow regions, except in the cases of H$_2$S and PH$_3$. For reasons identical to those invoked at the same epoch of PSN evolution in scenario 1, the surface density of solids is strongly decreased in the region centered at $\sim$7 au.

\begin{table}[h]
\centering
\caption{Heliocentric distance and value of the enrichment peak (relative to the initial abundance) for each species under consideration at $t$ = 0.1 Myr in the case of scenario 2 (middle column of Figure \ref{fig:am1-f}).}
\begin{tabular}{@{}ccc@{}}
\hline
\hline
Element         & Peak location (au)            & Peak value            \\ 
\hline
\ce{H2O}        & 2.7                                   & 8.5                   \\
\ce{CO}         & 4.8                                   & 7.0                    \\
\ce{CO_2}       & 4.7                                   & 7.5                   \\
\ce{CH_4}       & 4.7                                   & 7.0                   \\
\ce{H_2S}       & 4.7                                   & 7.1                   \\
\ce{N_2}        & 4.7                                   & 7.0                   \\
\ce{NH_3}       & 5.6                                   & 10.1                  \\
\ce{Ar}         & 4.7                                   & 7.0                   \\
\ce{Kr}         & 4.7                                   & 7.0                   \\
\ce{Xe}         & 4.7                                   & 7.0                   \\
\ce{PH_3}       & 4.7                                   & 7.1                   \\ 
\hline
\end{tabular}\\
Both vapor and solid forms are considered.
\label{tab:peaks_am}
\end{table}

\newpage

\section{Discussion}
\label{sec:sec4}

In this section, we discuss the implications of our model for various bodies of the solar system. We first investigate the sensitivity of our model to the variation of its input parameters and then provide fits of the volatile composition of comet R2 and those of Uranus and Neptune.

\subsection{Sensitivity to parameters}
\label{sec:Sensitivity to parameters}

The stability of our results has been tested against the variations of the pebble density, disk's mass, and the viscosity parameter in the 0.1--1 g cm$^{-2}$, $10^{-2}$--$10^{-1}$ $M_\odot$, and $10^{-4}$--$10^{-2}$ ranges, respectively. Variation in pebble density leads to results similar to those presented in Sec. \ref{sec:Scenario I}  and \ref{sec:Scenario II}. Lower density pebbles drift over shorter timescales at given size, but are also smaller because of a lower fragmentation limit (see Sect. \ref{sec2.2}). On the other hand, smaller pebbles drift over longer timescales, implying that both effects are (roughly) mutually counterbalanced and produce only minor variations in the abundance profiles. Both the  mass and $\alpha$ viscosity parameter of the disk affect its viscous evolution and, thus, the locations of the various icelines. Less massive disks are cooler because of their reduced viscous dissipation (see Eq. \ref{eq:temp}) and display their icelines closer to the host star. For example, we find that the condensation and clathration lines of the species investigated in our study are $\sim$2 au closer to the Sun in a $10^{-2}$ $M_\odot$ disk, compared with a $10^{-1}$ $M_\odot$ disk. As another example, the enrichment peak associated with the CO iceline ranges between 7 and 11 au from the Sun when the $\alpha$--value is varied between $10^{-4}$ and $10^{-2}$. The magnitude of the abundance peaks is also affected by the variation of $\alpha$. The abundance of the CO peak ranges between about 10 and 5 times, respectively, its initial PSN abundance when the $\alpha$ value is varied between $10^{-4}$ and $10^{-2}$. In our disk model, the contribution from the irradiation by the Sun to the disk's midplane temperature is not considered because it is assumed that the outer PSN is shadowed by its inner region \citep{Oh21}. This allows the disk to reach temperatures low enough to enable the condensation of ultravolatiles (CO, N$_2$, Ar, etc.) at the current locations of the giant planets, assuming the absence of migration during formation. When this contribution is included by considering the formalism depicted in \cite{ada88} and \cite{rud91}, the temperature profile becomes slightly warmer, implying that the icelines and clathration lines are moved outward by 1--2 au. Despite these changes, the general trend of our results is not impacted.

\subsection{Implications for comet R2}

R2 is a long-period comet displaying an unusually high N$_2$/CO ratio of 0.006--0.008 \citep{biv18,op19}. Another peculiar characteristic of this comet is its heavy depletion in terms of H$_2$O, with a CO/H$_2$O ratio of about 312 \citep{mck19}.

\begin{figure}[ht]
\center
\includegraphics[width=0.9\columnwidth, clip]{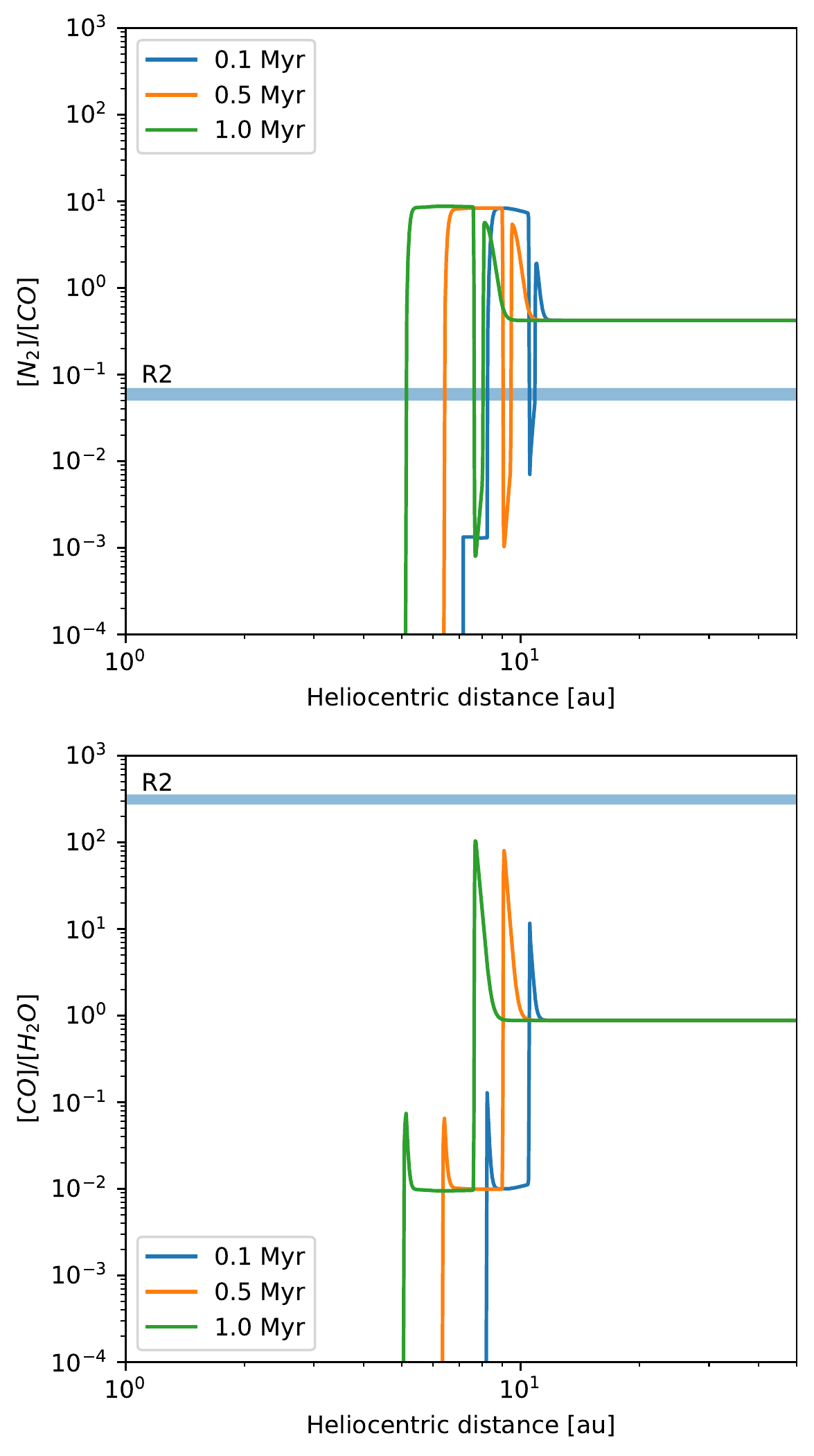}
\caption{N$_2$/CO (top panel) and CO/H$_2$O (bottom panel) abundance ratios in pebbles represented as a function of heliocentric distance in the case of scenario 1, and compared with those measured in R2's coma (blue bar) at 0.1, 0.5, and 1 Myr of the PSN evolution. Both ratios measured in R2 are simultaneously matched by our model at a distance of 7 au and 1 Myr.}
\label{fig:R2_ratios_cryst}
\end{figure}

\begin{figure}[ht]
\center
\includegraphics[width=0.9\columnwidth, clip]{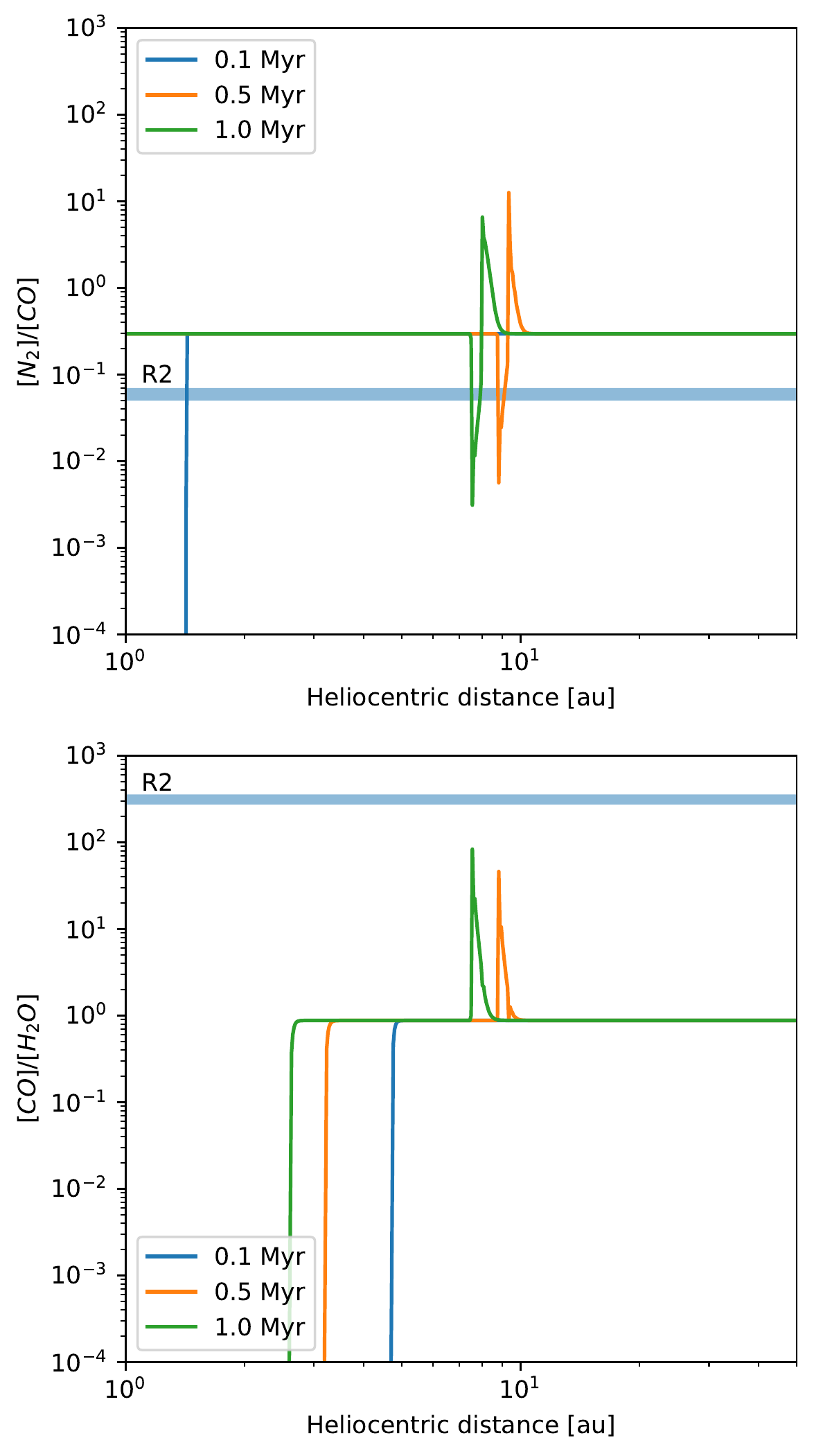}
\caption{N$_2$/CO (top panel) and CO/H$_2$O (bottom panel) abundance ratios in pebbles represented as a function of heliocentric distance in the case of scenario 2, and compared with those measured in R2's coma (blue bar) at 0.1, 0.5 and 1 Myr of the PSN evolution. Only the N$_2$/CO ratio is reproduced in R2 at 8 au and 1 Myr. The CO/H$_2$O ratio is however approached by our model at the same location and epoch of the PSN evolution.}
\label{fig:R2_ratios_am}
\end{figure}

Figures \ref{fig:R2_ratios_cryst} and \ref{fig:R2_ratios_am} represent the radial profiles of the CO/H$_2$O and N$_2$/CO ratios calculated in the PSN pebbles with our nominal model as a function of time in the cases of scenario 1 and scenario 2, respectively. Both ratios are compared with those measured in R2's coma (blue horizontal bar). Assuming that R2 formed from a unique set of building blocks, we require both ratios to be reproduced by our model at the same heliocentric distance and epoch. Simulations performed in the case of scenario 1 reproduce both ratios at a heliocentric distance of 7 au, after 1 Myr of PSN evolution. Our result is then consistent with those derived from the study of \cite{mou21b}, based on a simple approach that does not consider the interplay between the clathrate and water ice reservoirs.

Simulations performed in the case of scenario 2 reproduce the N$_2$/CO ratio at 8 au after 1 Myr of PSN evolution. However, the CO/H$_2$O ratio is not matched by our model, even if a peak is seen at 8 au and 1 Myr. We should note that the positions and magnitudes of those peaks can change when the $\alpha$--parameter and mass of our disk model are varied. This implies that, even if scenario 1 provides a better match of R2's composition with our nominal model, scenario 2 cannot be excluded. Interestingly, our calculated peaks are within a zone of dynamic instability in the early Solar System, which is more likely to result in ejection of planetesimals than capture by the Oort Cloud. This is a possible explanation for the lack of R2-like comets observed today \citep{An22}.

\subsection{Implications for the composition of Jupiter}
\label{sec:jupiter}

One-$\sigma$ error bar measurements made at Jupiter by the Galileo probe and the Juno spacecraft indicate C, N, O, S, P, Ar, Kr, and Xe abundances that are $\sim$1.5 to 6 times higher than the protosolar values \citep{at03,wo04,mou18,li20}. To explain those features, it has been proposed that Jupiter's atmosphere could reflect the composition of icy planetesimals either made of amorphous ice \citep{ow99} or from pure condensates and/or clathrates \citep{ga01,ga05,mou18,mou21b}. Alternatively, it has been proposed that this supersolar metallicity could result from the accretion of already pre-enriched PSN gas \citep{mou19,ag22}.

\begin{figure}[ht]
\centering
\includegraphics[width=0.9\columnwidth]{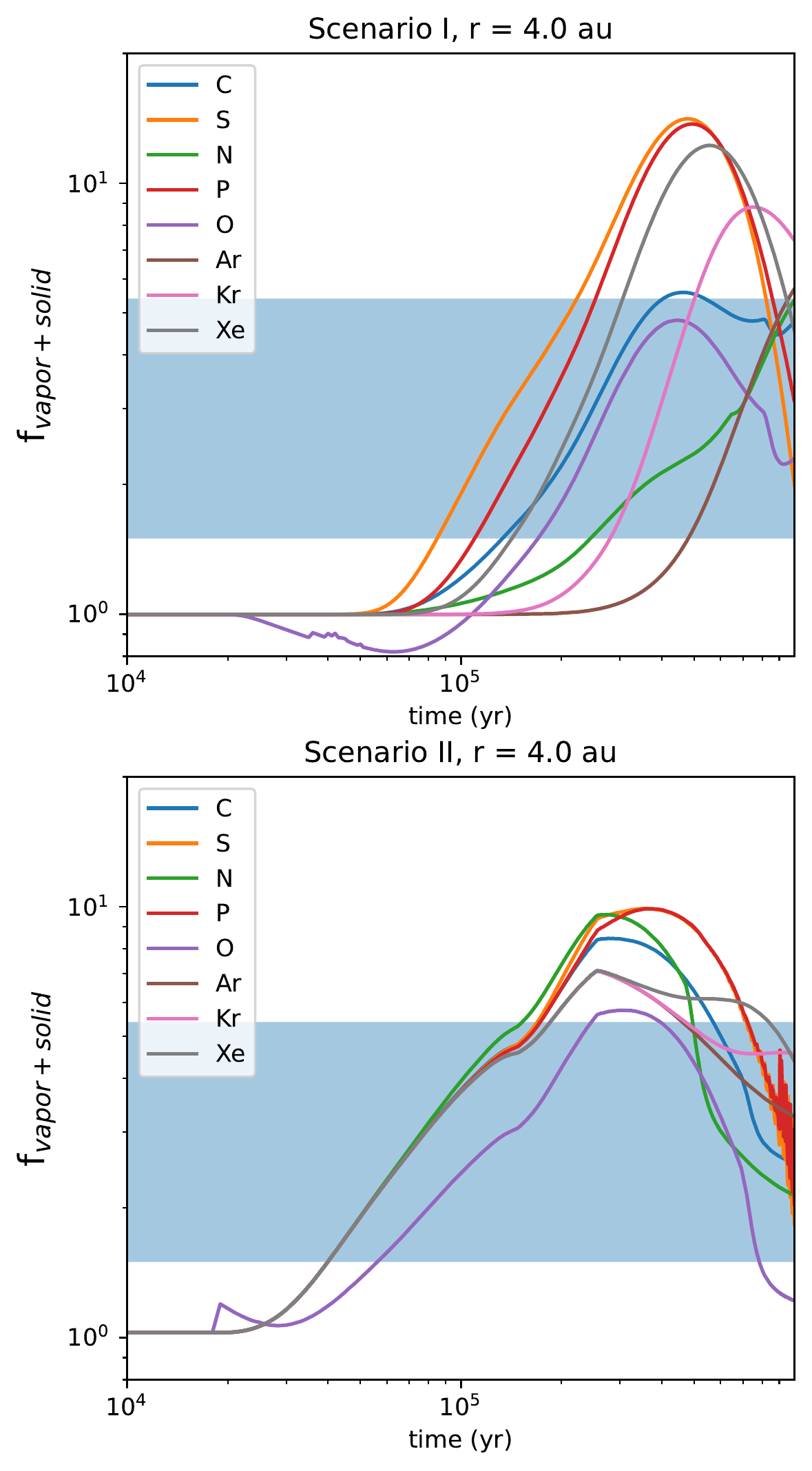}
\caption{Time evolution of the elemental abundances (relatives to the protosolar values) at a heliocentric distance of 4 au in the cases of scenario 1 (top panel) and scenario 2 (bottom panel). The blue area corresponds to the range covered by the elemental abundances derived from spacecraft measurements (see text). } 
\label{fig:jupiter}
\end{figure}

Figure \ref{fig:jupiter} represents the time evolution of the sum of the elemental enrichments calculated in vapor and solid phases at the heliocetric distance of 4 au, compared with their protosolar values, and in the cases of our two scenarios. Following the approach of \cite{ag22}, we focused on the composition of the PSN at 4 au, chosen as the location of Jupiter's formation. This distance is, in the model, beyond the water iceline but inward of the icelines of all other trace species. The dust-to-gas ratio can easily become greater than 2 to 3 times the protosolar composition in this region and could ease the formation of a proto-Jupiter core via the streaming instability \citep{ya17}.

Figure \ref{fig:jupiter} shows that the measured elemental enrichments are all matched by our model after 0.8--1 Myr and 50--100 kyr of the PSN evolution in scenario 1 and scenario 2, respectively. Our models suggest that the consideration of clathrate formation in addition to the crystallization of pure condensates in the PSN (scenario 1) still allow for the formation of a Jupiter-like planet from supersolar gases originating from the disk, compared with models considering the crystallization of pure condensates only, such as that developed by \cite{ag22}. Our model also suggests that the presence of multiple condensation and clathration lines does not alter the formation of supersolar vapors subsequent to their release from amorphous ice (scenario 2). Previous works exploring this possibility did not consider the formation of icelines in their models \citep{mon15,mou19}. 
In particular, Saturn's tropospheric abundances of C, N, S, and P have been measured to be between 3 and 13 times their protosolar abundances \citep{at18}, suggesting that its metallicity is higher than that of Jupiter. Assuming that Saturn formed in the vicinity of the CO$_2$ iceline to account for its high carbon enrichment, which corresponds to a heliocentric distance of $\sim$6 au at early epochs in our PSN model, our calculations show that this range of enrichments is reproduced within 0.2--0.3 Myr in scenario 1 and 0.2 Myr in scenario 2. This implies that Saturn could have formed earlier than Jupiter in scenario 1, whereas in scenario 2, it could have formed later. An earlier formation of Saturn could have reduced the inward flux of pebble and vapors, lowering the value of elemental enrichment peaks at Jupiter's location. However, the height of enrichment peaks is highly sensitive to the disk parameters (e.g., the $\alpha$-value; see \cite{ag22}). Therefore, if the accretion of pebbles by Saturn reduces the height of the enrichment peaks at Jupiter's location, it is still possible to achieve a metallicity similar to what is measured in its atmosphere.

\subsection{Implications for the composition of Ice Giants}
\label{sec:composition_of_ice_giants} 

\begin{figure}[ht]
\center
\includegraphics[width=0.9\columnwidth]{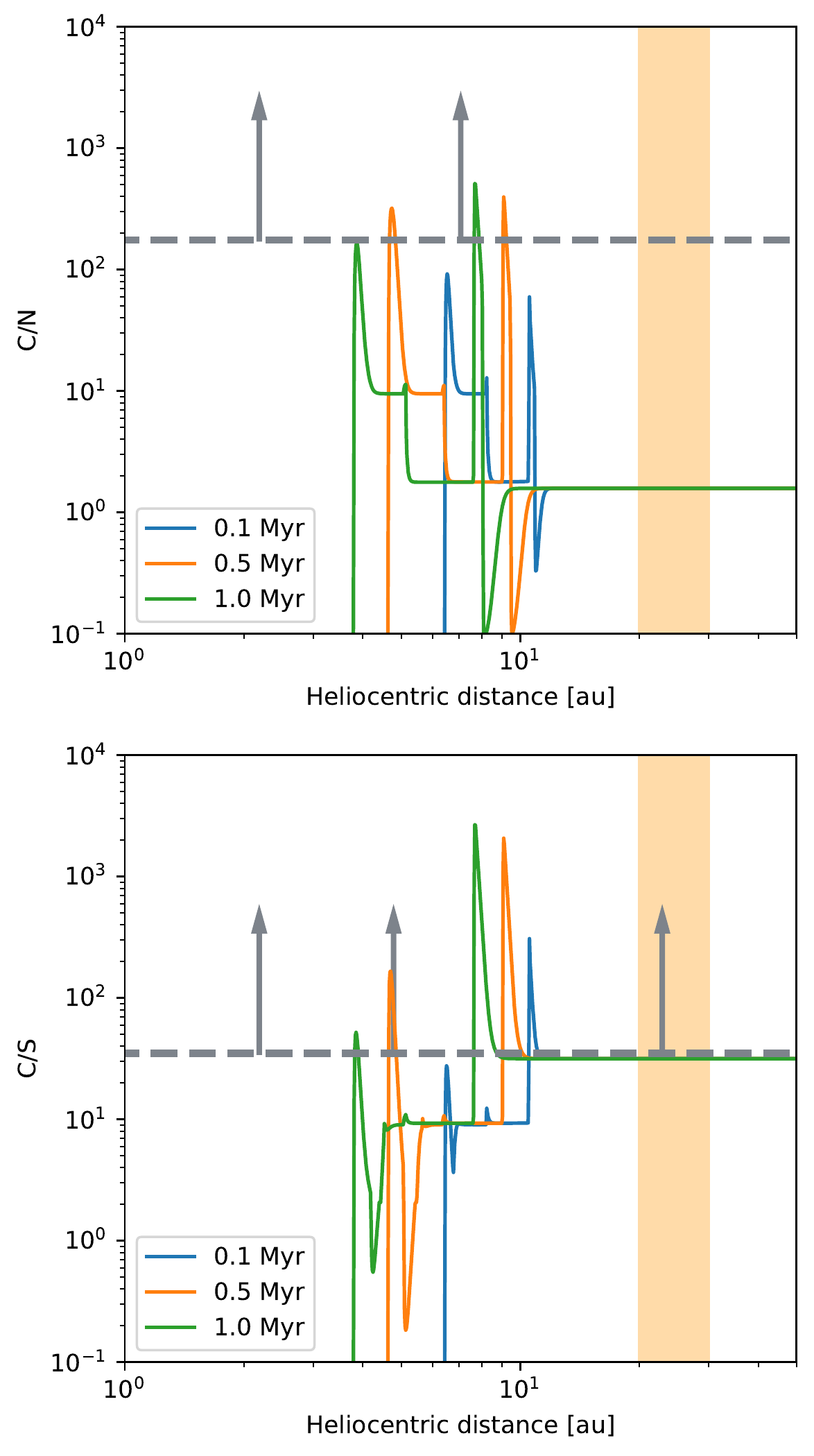}
\caption{Radial profiles of the C/N (top panel) and C/S (bottom panel) ratios calculated in pebbles at different epochs of the PSN evolution in the case of scenario 1. The horizontal dashed line represents the minimum ratio measured in the tropospheres of Uranus and Neptune. The orange area encompasses the current locations of the ice giants in the solar system.}
\label{fig:ice_giants_cryst}
\end{figure}

\begin{figure}[ht]
\center
\includegraphics[width=0.9\columnwidth]{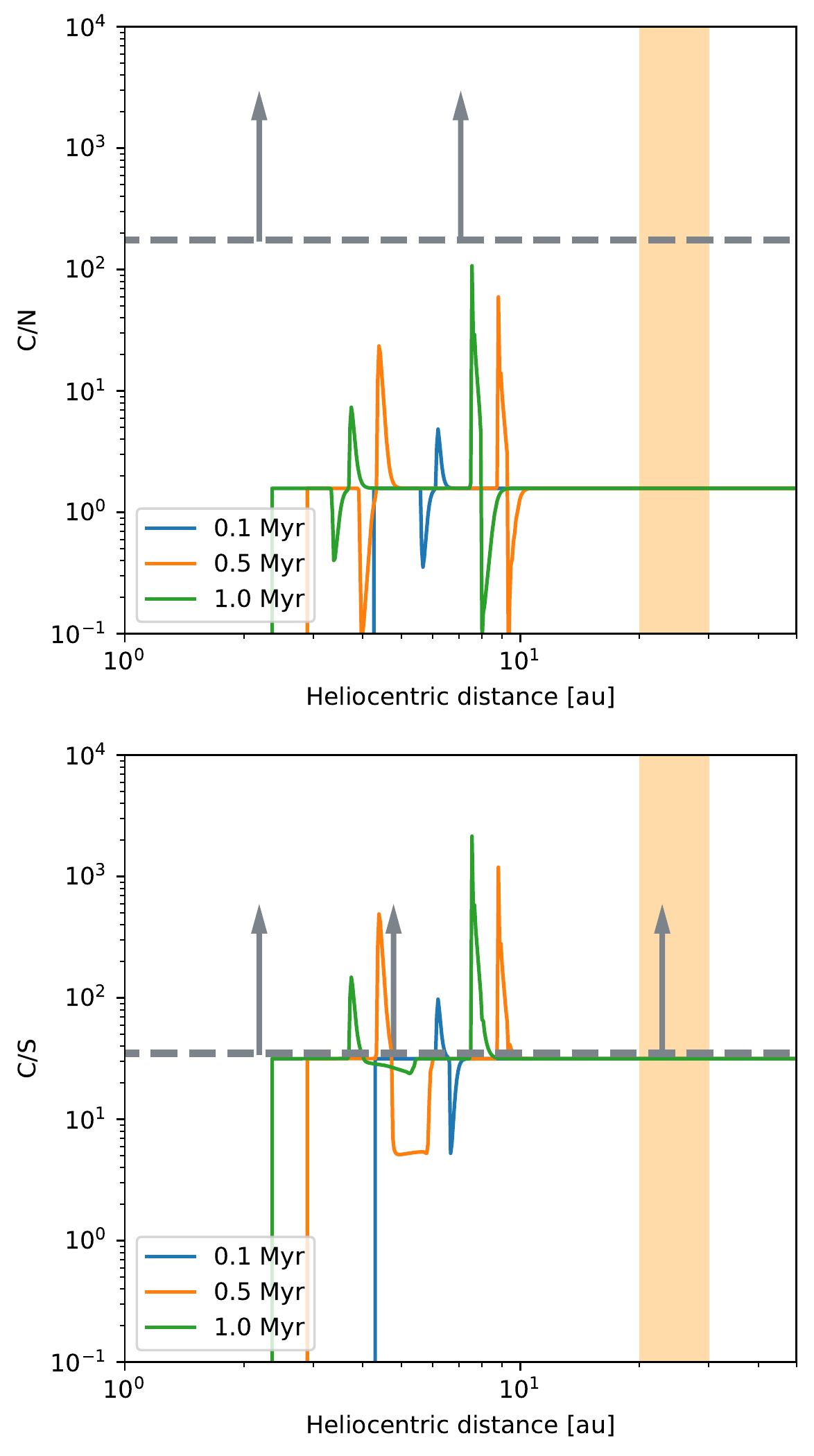}
\caption{Radial profiles of the C/N (top panel) and C/S (bottom panel) ratios calculated in pebbles at different epochs of the PSN evolution in the case of scenario 2. The horizontal dashed line represents the minimum ratio measured in the tropospheres of Uranus and Neptune. The orange area encompasses the current locations of the ice giants in the Solar System.}
\label{fig:ice_giants_am}
\end{figure}

The composition of the deep atmospheres of Uranus and Neptune is shrouded in mystery since most of the heavy constituents condense at pressures deeper than may readily be probed remotely \citep{mo20}. The only determinations that can be used so far in our model are the C/N and C/S ratios, which have been found equal to or higher than $\sim$175 and $\sim$35, respectively \citep{as09,ka09,ka11,ir18,ir19a,ir19b}. Figures \ref{fig:ice_giants_cryst} and \ref{fig:ice_giants_am} represent the radial profiles of the  C/N and C/S elemental ratios in pebbles as a function of time in the PSN, compared with the minimum values measured in the tropospheres of the two ice giants. In both cases, these ratios can be reproduced in the $\sim$7--8 au region after 1 Myr of PSN evolution. The formation distance of solids in the PSN is model-dependent, but our simulations suggest that the giant planets did form in a more compact configuration than current the one -- which is also in agreement with several dynamical models  \citep{ts05,lyk10,gu11}. 
Given their high metallicities, the formation of Uranus and Neptune requires a higher surface density of solids than the values derived here for the outer PSN. This assumption is at odds with the fact that the planetesimal density and collision probability are both low in the outer disk.
One way to overcome this difficulty would be to assume the formation of the two giants via the accretion of pebbles directly onto the planetary embryo, which can still work efficiently far from the host star \citep{hel14,bit15,bit18,ar19}.
 
\subsection{Limitations of the model} 

Over the recent years, substructures have been largely observed in protoplanetary disks. The ALMA/DSHARP survey showed that protoplanetary disks are not smooth and that substructures are ubiquitous \citep{an20,jen22}. Substructures can be produces by planet-disk interactions, in the form of spiral density waves \citep{mu12,zha11} and gaps \citep{bae17}. Radiative hydrodynamic models show that these substructures can be also formed by instabilities in the disk \citep{lov99,lov14,bla21}. Such features translate into local density and pressure variations that act as dust traps which locally enhance the dust surface density. Although our model does not take into account such disk substructures, we show that the icelines, clathration lines, and the ACTZ correspond to vapor and pebble enrichment peaks, which can lead to instabilities.

In this work, giant planets are assumed to be formed in situ. In scenario 1, Jupiter is formed within 0.8--1 Myr. This timescale is compatible with a formation via pebble accretion \citep{bit15,bit18,ali18,ar19,ven20} and gravitational instability \citep{bos97,zhu12,kra16}. On the other hand, in scenario 2, Jupiter forms in less than 0.1 Myr. This timescale is consistent with a gravitational instability which can be triggered as early as 0.1 Myr \citep{zhu12}. An early formation of Jupiter allows us to account for the observed carbonaceous chondrites dichotomy \citep{kle20}. Although, the disk instability model is consistent with formation timescales found in both scenario 1 and scenario 2, it is important to note that observations and models suggest that, at minimum, amorphous ice should be heated up to its crystallization temperature when falling from the presolar cloud onto the PSN \citep{vi13}. Those findings suggest that scenario 1 is the most likely scenario, implying that our results remain consistent with the core accretion model.

One limitation of our model is the fact that planet migration is not considered. During or after formation, planets migrate inward or outward \citep{mas03,bit15,sch21}. During migration, planets accrete material from different parts of the disk with various compositions. Although planetary migration contradicts an in situ formation hypothesis, 3D hydrodynamical simulations indicate that the rates of type I and type II migrations could be several times slower than the prescription usually employed in the literature of planet formation \citep{chr20,leg21,cha22}. Assuming an in situ formation of Jupiter in our model to reproduce its observed metallicity is expected to remain valid in light of these revised migration rates because our derived timescales are still quite short, compared to the PSN evolution.

\section{Conclusion}
\label{sec:sec5}

In this work, we investigate the impact of clathrate formation on the radial distribution of volatiles in the PSN, considering two scenarios, each of them corresponding to a distinct initial condition. To do so, we used a 1D protoplanetary disk model coupled with modules describing the evolution of trace species in the vapor phase, as well as the dynamics of dust and pebbles. This model also considers the different sources and sinks for the volatile phases considered (vapors, pure condensates, or clathrates). 

In scenario 1, we assume that the volatiles were delivered to the PSN in the form of pure condensate grains. In this case, we show that clathrates can crystallize and form enrichment peaks at about 10 times the value of the initial abundances at their clathration lines, which are closer to the Sun than their corresponding icelines. The amount of clathrates formed in the PSN depends on the local abundance of crystalline water, which in many cases, acts as a limiting factor in our model. In scenario 2, we assumed that the volatiles were delivered to the PSN in the form of amorphous grains. Under those conditions, volatiles are only released from amorphous ice when the icy grains are heated up to $\sim$135K, namely, at the ACTZ location. An enrichment peak up to about seven times the initial abundances then forms at the ACTZ location. In this case, clathrate formation is not possible because there is no crystalline water ice available beyond the ACTZ in the PSN. All the enrichment peaks of pure condensates are also located close to the ACTZ. Our investigation shows that both scenario 1 and scenario 2 can reproduce the known compositions of comet R2, Jupiter, Uranus, and Neptune. This implies that our model does not allow us to formally rule out the presence of amorphous ice during the early phases of the PSN, assuming that those planetary bodies accreted from pebbles or pebble-made planetesimals. More planetary composition data, such as the in situ measurements of Saturn's atmosphere, are needed to understand the initial formation conditions of the PSN.

\section{Acknowledgments}

The project leading to this publication has received funding from the Excellence Initiative of Aix-Marseille Universit\'e -- A*Midex, a French ''Investissements d'Avenir programme'' AMX-21-IET-018. This research holds as part of the project FACOM (ANR-22-CE49-0005-01\_ACT) and has benefited from a funding provided by l’Agence Nationale de la Recherche (ANR) under the Generic Call for Proposals 2022. OM acknowledges support from CNES. JIL was supported by the Juno project. We acknowledge Sarah E. Anderson for helpful discussions about her dynamical simulations regarding the origin of comet R2. We thank the anonymous referee for their useful comments that improved the quality of this paper.

\appendix

\section{Vapor pressures of pure condensates}
\label{an:condensates}

The vapor pressures of the pure condensates considered in this work, except H$_2$O and NH$_3$, follow the law:

\begin{equation}
\ln{P_{\mathrm{eq}}} = \sum_k a_k \left(\frac{1}{T}\right)^k
,\end{equation}

\noindent where the $a_k$ factors have been determined experimentally \citep{fr09} and are summarized in Table \ref{tab:eq_pure}. 

The vapor pressures of H$_2$O is given by \citep{wa11}:

\begin{equation}
\begin{split}
P_{\mathrm{eq}} = & P_{\mathrm{tp}} \exp{\left( \frac{T_{\mathrm{tp}}}{T} \right)} \left( -21.2144006 \left(\frac{T_{\mathrm{tp}}}{T}\right)^{1/300} \right. \\
&   +27.3203819 \left(\frac{T_{\mathrm{tp}}}{T}\right)^{2.10666667}  \\
& \left. -6.10598130 \left(\frac{T_{\mathrm{tp}}}{T}\right)^{1.70333333} \right),
\end{split}
\end{equation}

\noindent where $P_{\mathrm{tp}}$ and $T_{\mathrm{tp}}$ are the pressure and the temperature of the triple point of water, respectively. 

The phosphine equilibrium pressure is given by \cite{st47}:
\begin{equation}
\log_{10}{ P_{\mathrm{eq}}} = 4.02591 - \frac{702.651}{ T - 11.065}
\label{eq:ph3}
.\end{equation}

\begin{table*}[ht]
\centering
\caption{Polynomial factors for the vapor pressure equations of pure condensates}
\begin{tabular}{lccccccccc}
\hline
\hline\\
Element & Temperature range (K) & $a_0$ & $a_1$ & $a_2$ & $a_3$ & $a_4$ & $a_5$ & $a_6$\\
\hline
$\ce{CO}$ & T $\le 61.55$ & 10.43 & -721.3 & -10740 & $2.341 \times 10^5$ & $-2.392 \times 10^6$ & $9.478 \times 10^6$ & $\emptyset$ \\
&  T $> 61.55$ & 10.25 & -748.2 & -5843 & $3.939 \times 10^4$ & $\emptyset$ & $\emptyset$ & $\emptyset$ \\
\ce{CO2} & T $\le 40$ & $10^{-40}$ & $\emptyset$ & $\emptyset$ & $\emptyset$ & $\emptyset$ & $\emptyset$ & $\emptyset$ \\ 
&  $40.0 <$ T $\le 194.7$ & 14.76 & -2571 & $-7.781 \times 10^{4}$ & $4.325 \times 10^6$ & $-1.207 \times 10^8$ & $1.350 \times 10^9$ & $\emptyset$ \\
&  T $> 194.7$ & 18.61 & -4154 & $1.041 \times 10^5$ & $\emptyset$ & $\emptyset$ & $\emptyset$ & $\emptyset$\\
\ce{CH4} & all & 10.51 & -1110 & -4341 & $1.035 \times 10^5$ & $ -7.910 \times 10^5$ & $\emptyset$ & $\emptyset$  \\ 
\ce{H2S} & T $\le 127.0$ & 12.98 & -2707 & $\emptyset$ & $\emptyset$ & $\emptyset$ & $\emptyset$ & $\emptyset$ \\           
& T $> 127.0$ & 8.933 & -726.0 & $-3.504 \times 10^5$ & $2.724 \times 10^7$ & $-8.582 \times 10^8$ & $\emptyset$ & $\emptyset$ \\
\ce{N2} & T $\le 35.61$ & 12.40 & -80.74 & -3926 & $6.297 \times 10^4$ & $-4.633 \times 10^5$ & $1.325 \times 10^5$ & $\emptyset$ \\ 
& T $> 35.61$ &  8.514 & -456.4 & $-1.987 \times 10^4$ & $4.800 \times 10^5$ & $-4.524 \times 10^6$ & $\emptyset$ & $\emptyset$  \\
\ce{NH3} & all & 15.96 & -3537 & $-3.310 \times 10^4$ & $1.742 \times 10^6$ & $-2.995 \times 10^7$ & $\emptyset$ & $\emptyset$ \\ 
\ce{Ar} & all & 10.69 & -893.2 & -3567 & $6.574 \times 10^4$ & $-4.280 \times 10^5$ & $\emptyset$ & $\emptyset$ \\ 
\ce{Kr} & all & 10.77 & -1223 & -8903 & $2.635 \times 10^5$ & $-4.260 \times 10^6$ & $3.575 \times 10^7$ & $-1.210 \times 10^8$ \\ 
\ce{Xe} & all & 10.698 & -1737 & $-1.332 \times 10^4$ & $4.349 \times 10^5$ & $-7.027 \times 10^6$ & $4.447 \times 10^7 $ & $\emptyset$ \\ 
\hline
\end{tabular}
\label{tab:eq_pure}
\end{table*}

\section{Dissociation pressures of NH$_3$ monohydrate and clathrates}
\label{an:clathrate}

The dissociation pressures of NH$_3$ monohydrate and clathrates follow the Antoine law: 

\begin{equation}
\ln{P_{\mathrm{eq}}} = \frac{A}{T} + B,
\end{equation}

\noindent with $A$ and $B$ parameters determined from experiments and summarized in Table \ref{tab:peq_cla}.

\begin{table}[h]
    \centering
    \caption{Parameters for the dissociation pressure equations of NH$_3$ monohydrate and clathrates}
    \begin{tabular}{lccc}
        \hline
        \hline
        Element & A & B & Reference \\
        \hline
        \ce{CO} & -1685.54 & 10.9946 & \cite{he04} \\
        \ce{CO2} & -2544.395 & 11.411518 & \cite{lon05} \\ 
        \ce{CH4} & -2161.81 & 11.1249 & \cite{he04} \\ 
        \ce{H2S} & -3111.02 & 11.3801 & \cite{he04} \\ 
        \ce{N2} & -1677.62 & 11.1919 & \cite{he04} \\ 
        \ce{NH3} & -2878.28 & 8.00205 & \cite{he04} \\ 
        \ce{Ar} & -1481.78 & 9.95523 & \cite{he04} \\ 
        \ce{Kr} & -1987.5 & 9.99046 & \cite{he04} \\ 
        \ce{Xe} & -2899.19 & 11.0354 & \cite{he04} \\ 
        \ce{PH3} & -3011.28 & 11.95 & \cite{lu85}\\
        \hline
    \end{tabular}
    \label{tab:peq_cla}
\end{table}

\end{document}